\documentclass[structabstract]{aa}  
\usepackage{graphicx}
\usepackage{txfonts}
\usepackage{longtable,lscape}
\usepackage{natbib}
\bibpunct{(}{)}{;}{a}{}{,}
\begin{document}

\title{The low-mass population of the $\rho$~Ophiuchi molecular cloud 
\thanks{Based on observations obtained with WIRCam, a joint project of CFHT, Taiwan, Korea, Canada, France, at the Canada-France-Hawaii Telescope (CFHT) which is operated by the National Research Council (NRC) of Canada, the Institute National des Sciences de l'Univers of the Centre National de la Recherche Scientifique of France, and the University of Hawaii. Based on observations made at the ESO La Silla and Paranal Observatory under program 083.C-0092. Based in part on data collected at Subaru Telescope, and obtained from the SMOKA, which is operated by the Astronomy Data Center, National Astronomical Observatory of Japan. Research supported by the Marie Curie Research Training Network CONSTELLATION under grant no. MRTN-CT- 2006-035890.}}

\author{C. Alves de Oliveira\inst{1}, E. Moraux\inst{1}, J. Bouvier\inst{1}, H. Bouy\inst{2}, C. Marmo\inst{3}, L. Albert\inst{4} }
\institute{Laboratoire d$^{\prime}$Astrophysique de Grenoble, Observatoire de Grenoble, BP 53, 38041 Grenoble Cedex 9, France \\
\email{Catarina.Oliveira@obs.ujf-grenoble.fr}
\and
Herschel Science Centre, European Space Agency (ESAC), P.O. Box 78, E-28691 Villanueva de la Ca\~{n}ada, Madrid, Spain
\and
Institut d$^{\prime}$Astrophysique de Paris, 98bis Bd Arago, 75014 Paris, France 
\and
Canada-France-Hawaii Telescope Corporation, 65-1238 Mamalahoa Highway, Kamuela, HI 96743, USA }
\date{Received December 18 2009; accepted February 27 2010}


\abstract
{Star formation theories are currently divergent regarding the fundamental physical processes that dominate the substellar regime. Observations of nearby young open clusters allow the brown dwarf (BD) population to be characterised down to the planetary mass regime, which ultimately must be accommodated by a successful theory.}
{We hope to uncover the low-mass population of the $\rho$ Ophiuchi molecular cloud and investigate the properties of the newly found brown dwarfs.}
{We use near-IR deep images (reaching completeness limits of approximately 20.5~mag in \emph{J}, and 18.9~mag in \emph{H} and  \emph{K$_{s}$}) taken with the Wide Field IR Camera (WIRCam) at the Canada France Hawaii Telescope (CFHT) to identify candidate members of $\rho$ Oph in the substellar regime. A spectroscopic follow-up of a small sample of the candidates allows us to assess their spectral type, and subsequently their temperature and membership.}
{We select 110 candidate members of the $\rho$ Ophiuchi molecular cloud, from which 80 have not previously been associated with the cloud. We observed a small sample of these and spectroscopically confirm six new brown dwarfs with spectral types ranging from M6.5 to M8.25.}
{}

\titlerunning{The low-mass population of the $\rho$~Ophiuchi molecular cloud} 
\authorrunning{C. Alves de Oliveira et al.} 
\maketitle


\section{Introduction}
\label{introduction}

The determination of the initial mass function (IMF) across the entire stellar and substellar mass spectrum is a fundamental constraint for star formation theories \citep[see, for example,][and references therein]{Bonnell2007}. Although there are general accepted views on the way star formation occurs and young stellar objects (YSOs) evolve to the main sequence \citep{Shu1987,Larson1973}, the existing theories have not yet converged to an agreed paradigm that can explain the wide range of existing observational properties of YSOs. In particular, since their discovery, hundreds of brown dwarfs (BD) with masses down to the planetary regime have been uncovered in star-forming regions and the solar neighbourhood, with a ratio of the number of BDs to stars of approximately 1/5 \citep[see, for example,][ and references therein]{Luhman2007c}, implying that a successful star and planet formation theory must account for them. Different theories for the formation of BDs are currently debated, according to which they could either form by gravitational fragmentation and collapse of molecular cores \citep{Padoan2007,Hennebelle2008}, from early ejection from stellar embryos \citep{Reipurth2001,Whitworth2005}, or from fragmentation of massive circumstellar discs \citep{Stamatellos2009}. The extension of the IMF to the brown dwarf and planetary mass regime and the search for the end of the mass function is therefore crucial to determine the dominant formation process of substellar objects and its relation with the surrounding environment \citep{Moraux2007,Andersen2008,Luhman2007}. Brown dwarfs are brighter when they are young \citep{Chabrier2000} and their detection down to a few Jupiter masses can be attained with the current technology by studying them in young star-forming regions \citep{Lucas2000,ZapateroOsorio2002,Weights2009,Burgess2009,Marsh2009}. For that reason, one of the prime goals of modern observations is to achieve completeness at the lower mass end, i.e., the brown dwarf and planetary mass regime, for different environments across several young star-forming regions \citep[][among many others]{Bihain2009,Bouy2009b,Bouy2009a,Lodieu2009,Luhman2009,Scholz2009}. 

The main motivation of our survey of the $\rho$~Ophiuchi molecular cloud is to uncover the low-mass population of the cluster down to the planetary regime. Despite being one of the youngest ($\sim$1~Myr) and closest star-forming regions \citep[120 to 145~pc,][]{Lombardi2008,Mamajek2008},  the high visual extinction in the cloud's core, with A$_V$ up to 50-100 mag \citep{Wilking1983}, make it one of the most challenging environments to study low-mass YSOs. The main studies previously conducted in $\rho$~Oph have been summarised in a recent review \citep{Wilking2008}, which includes a census with the $\sim$300 stellar members that have been associated with the cloud up to now, from which only 15 are estimated to have masses in the substellar regime. \citet{Marsh2009} reported the discovery a young brown dwarf with an estimated mass of $\sim$2~$-$3~Jupiter masses in $\rho$~Oph, although we here question its membership to the cloud (see Sect. ~\ref{comp:surveys}).

We conducted a deep near-IR (\emph{J}, \emph{H}, and \emph{K$_{s}$}) photometric survey centred approximately on the cloud's core and covering $\sim$1~deg$^{2}$, which we use to identify candidate members in the substellar mass regime. Near-IR surveys are particularly suitable to study this star-forming region because most of its population is visibly obscured. Previous near-IR studies of this cluster have been done from the ground down to a sensitivity limit of \emph{K}~$<$~13-14~mag for a larger area of the cloud \citep{Greene1992,Strom1995,Barsony1997}, and of \emph{K}~$<$~15.5~mag for a smaller region \citep[200~arcmin$^2$,][]{Comeron1993}. Deeper observations were done from space with a small coverage of 72~arcmin$^2$ and a sensitivity of \emph{H}~$<$~21.5~mag \citep{Allen2002}. The WIRCam near-IR survey presented takes advantage of a new generation of wide-field imagers on 4 meter-class telescopes, to reach completeness limits of approximately 20.5 in \emph{J}, and 18.9 in \emph{H} and  \emph{K$_{s}$} over the entire degree-size area of the sky occupied by the $\rho$~Ophiuchi central cloud. This work complements the previous surveys both in the area it covers and in sensitivity. Of comparable characteristics is the near-IR survey recently conducted by \citet{Alvesdeoliveira2008}, which uses a different technique, near-IR variability, to select candidate members. Our selection method allows BDs with masses down to a few Jupiter masses (according to evolutionary models) to be detected through $\sim$20 magnitudes of extinction. Extensive use of archive data at optical and IR wavelengths is made to further characterise the candidate members. In a pilot study, a spectroscopic follow-up of a subsample of these candidates has confirmed six new brown dwarfs.

In Sects.~\ref{data} and \ref{archdata}, the observations and reductions for new and archive data are described. Section~\ref{select:cmd} explains the methods used to select candidate members of $\rho$~Oph and the results, and in Sect.~\ref{discussion_phot} we discuss their properties. Section~\ref{spec} describes the numerical fitting procedure used to analyse the data from the spectroscopic follow-up and the spectral classification. These results are then discussed through Sect. ~\ref{properties}. Conclusions are given in Sect. ~\ref{conclusion}.


\section{Observations and data reduction}
\label{data}

   \begin{table*}
    \begin{minipage}[t]{\linewidth}
   \caption{Journal of the WIRCam/CFHT observations.}         
   \centering             
    \renewcommand{\footnoterule}{}
       \begin{tabular}{l l l l l}       
   \hline           
   \hline
   Pointing & RA & Dec. & Date & Filters\footnote{For the \emph{J} filter, two sets of images were taken: short (7x4x5~s) and long (7x8x27~s) exposures. } \\
   ~ & (J2000) & (J2000) & ~ & ~ \\
   \hline                        
   CFHTWIR-Oph-A 	& 16 27 10.0 	& $-24$ 26 00	&  19 April 2006 	& \emph{J}, \emph{H}, \emph{K$_{s}$}  \\ 
   CFHTWIR-Oph-B 	& 16 25 50.0 	& $-24$ 26 00 	&  20 April 2006   	& \emph{J}, \emph{H}, \emph{K$_{s}$} \\
   CFHTWIR-Oph-C 	& 16 27 10.0 	& $-24$ 44 00 	&  09 May 2006   	& \emph{J}, \emph{H}, \emph{K$_{s}$} \\
   CFHTWIR-Oph-D 	& 16 28 29.0 	& $-24$ 44 00 	&  11 May 2006  	& \emph{J}, \emph{H}, \emph{K$_{s}$} \\
   CFHTWIR-Oph-E 	& 16 25 50.0 	& $-24$ 08 00 	&  12 May 2006   	& \emph{J},\emph{H}, \emph{K$_{s}$}   \\
   CFHTWIR-Oph-F 	& 16 27 10.0 	& $-24$ 08 00 	&  17 May 2006   	& \emph{J}, \emph{H}, \emph{K$_{s}$} \\
   CFHTWIR-Oph-G 	& 16 28 29.0 	& $-24$ 26 00	&  11 July 2006   	& \emph{J}, \emph{H}, \emph{K$_{s}$}  \\
    			&		     	&		        	&  17 July 2006		& \emph{J}		\\
   \hline                                 
   \label{table:1}   
   \end{tabular}
   \end{minipage}
   \end{table*}

 \begin{figure} 
 \centering
 \includegraphics[width=\columnwidth]{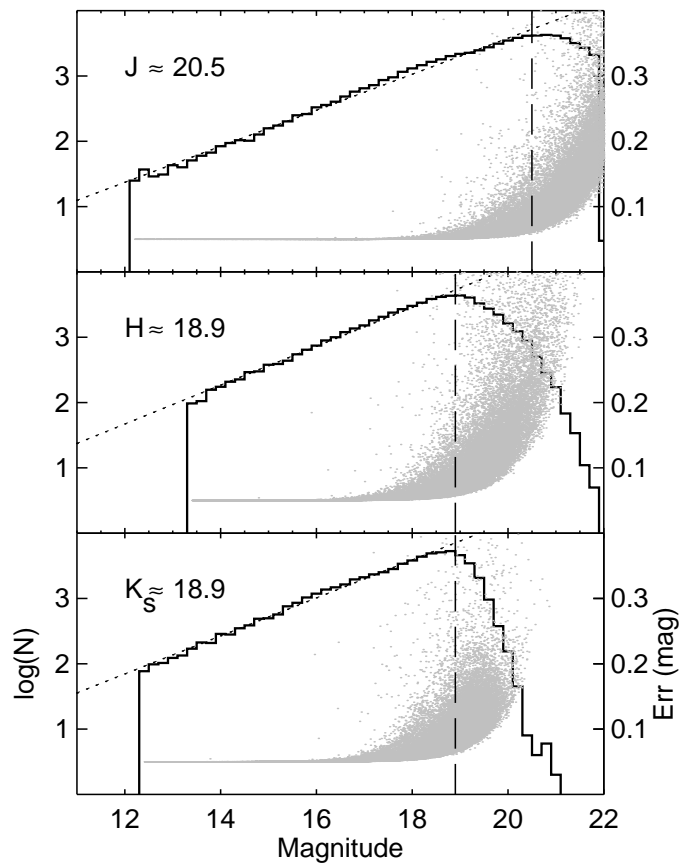}
   \caption{Histogram of the number of objects detected per magnitude bin and the respective magnitude errors. The points where the histograms diverge from a linear fit to the logarithmic number of objects per magnitude bin give an approximation of the completeness limit of the survey for the different filters: 20.5 in \emph{J}, and 18.9 in \emph{H} and  \emph{K$_{s}$}, with errors below $\sim$0.1 magnitudes.}
    \label{histmag}
    \end{figure}

We present the deep infrared photometric survey we conducted in the $\rho$~Ophiuchi cluster in the \emph{J}, \emph{H}, and \emph{K$_{s}$} filters, and the infrared spectroscopic follow-up of a small sample of candidate members of this star-forming region.

\subsection{The WIRCam/CFHT near-IR survey}
\label{data:wircam}
The WIRCam at the CFHT telescope is a wide-field imaging camera operating in the near-infrared, consisting of four Hawaii-II2-RG 2048~x~2048 array detectors with a pixel scale of 0.3$\arcsec$ \citep{Puget2004} . The four detectors are arranged in a 2~x~2 pattern, with a total field of view of 20$\arcmin$~x~20$\arcmin$. 

The data were obtained in queue-scheduled observing mode over several runs as part of a large CFHT key programme aimed at the characterisation of the low-mass population of several young star-forming regions (P.I. J. Bouvier). Seven different WIRCam pointings were needed to cover the central part of the $\rho$~Ophiuchi cluster. These were taken at different epochs over a three-month period, but all observations were done under photometric conditions, with a seeing better than 0$\farcs$8 (measured in the images to be typically between 0$\farcs$4 and 0$\farcs$5), and an airmass less than 1.2. All individual tiles were observed in the \emph{J}, \emph{H}, and \emph{K$_{s}$} filters, using a seven point dithering pattern selected to fill the gaps between detectors, and accurately subtract the sky background. Table~\ref{table:1} shows the central position (right ascension and declination) for each of the seven tiles and the dates of the observations. For each field, short and long exposures were obtained with the \emph{J} filter (7x4x5~s and 7x8x27~s, respectively), and shorter individual exposures with the \emph{H} and \emph{K$_{s}$} filters (7x8x7~s). 

Individual images are primarily processed by the 'I'iwi reductions pipeline at the CFHT (Albert et al., \emph{in prep.}), which includes detrending (e.g. bias subtraction, flat-fielding, non-linearity correction, cross-talk removal), sky subtraction, and astrometric calibration. Afterwards, the data are handled by Terapix \citep{Marmo2007}, the data reduction centre at the \emph{Institut d'Astrophysique de Paris} (France) responsible for carrying out the final quality assessment of the individual images,  determining precise astrometric and photometric calibrations, and combining the dither and individual exposures into the final stacked images. All images are merged into a single tile of $\sim$1~deg$^{2}$ centred on the cloud's core. The photometric calibration of the WIRCam data is done with 2MASS stars in the observed frames as part of the nominal pipeline reduction. Typical estimated errors in the WIRCam zero point determination are of $\sim$0.05~mag. Ultimately the photometric accuracy for a determined field is also dependent on the number of 2MASS stars available (which is probably reduced in regions in the sky with high extinction as is the case of some regions of $\rho$~Oph), if the stars used are themselves variable \citep{Alvesdeoliveira2008}, and lastly, photometric offsets can also occure from small problems in flatfielding and/or sky subtraction from night to night, given that the data were taken in different epochs.

We extracted PSF photometry from the mosaicked images with PSFEx (PSF Extractor, Bertin et al., \emph{in prep.}), a software tool that computes a PSF model from well-defined stellar profiles in the image, which is given as an input to the SExtractor programme \citep{Bertin1996} to compute the photometry for each detected object. During this first stage of the analysis, and to ensure the detection of all the faint sources present in the images, the extraction criteria used are not too stringent. An object is extracted if it complies with the required minimum to have three contiguous pixels with fluxes 1.5~$\sigma$ above the estimated background. An inspection of the images and detections showed that the number of spurious detections is minimum, while all the objects seen \emph{by eye} are detected. Catalogues of the short and long exposures for the \emph{J} filter are merged into one single catalogue. The overlap in magnitudes between the short and long exposures allows checking of the photometric accuracy. For objects that are common to the two catalogues, the two magnitude values for each object are compared, and  the r.m.s. accuracy is measured to be below 0.05 magnitudes.  The histogram of the magnitudes (not corrected for extinction) is shown in Fig.~\ref{histmag}. We derived approximate completeness limits of 20.5 in \emph{J}, and 18.9 in \emph{H} and \emph{K$_{s}$}, with errors below $\sim$0.1 magnitudes, located at the points where the histograms diverge from the dotted lines \citep{Wainscoat1992,Santiago1996}, which represent a linear fit to the logarithmic number of objects per magnitude bin, calculated over the intervals of better photometric accuracy. The catalogues from all filters are combined into a single database by requiring a positional match better than 1$\arcsec$ and detections across the \emph{J},  \emph{H}, and \emph{K$_{s}$} lists. The mean separation for the $\sim$27,000 detections common to the \emph{J} band short and long catalogues is found to be $\sim$0$\farcs$05, and when combining all the bands is $\sim$0$\farcs$1. The final catalogue contains $\sim$57,000 objects. Approximately 1000 of the brightest stars have a counterpart in the 2MASS catalogues, and the mean magnitude differences between the two systems is found to be 0.05, 0.07, and 0.09~mag for \emph{J},  \emph{H}, and \emph{K$_{s}$} respectively, which is of the order of expected zero point uncertainties. The dispersion of the differences can however be as high as $\sim$0.1~mag, which reflects the possible sources of error mentioned, and therefore we did not correct the photometry for these offsets. Furthermore, the WIRCam and 2MASS filters design differ substantially, and at no point are these colour effects taken into account. Throughout this work, all the WIRCam \emph{J},  \emph{H}, and \emph{K$_{s}$} photometry is  given in the CFHT Vega system.

\subsection{The SofI / NTT spectroscopic follow-up}
\label{data:ntt}
A spectroscopic follow-up was conducted for a subsample of 13 candidate members of the $\rho$~Ophiuchi Cluster, with magnitudes ranging from 12.5 to 15 in  \emph{K$_{s}$}. The selection of  candidate members is described in Sect.~\ref{select:cmd}. We also observed GY~201, which was previously associated with the cloud based on its mid-IR colours \citep{Wilking2008} but was not selected with our criteria. All observations were gathered from 3-6 May 2009, using SofI (Son of ISAAC), a near-IR low resolution spectrograph mounted on the 3.6~m New Technology Telescope (NTT, La Silla, ESO). The majority of the targets were observed with the blue and red grisms, which operate from 0.95 to 1.64, and 1.53 to 2.52~$\mu$m, respectively. Some objects were too faint in the \emph{J} band and could only be observed with the red grism. In addition, we observed nine sources that are not part of the candidate member list, but have magnitudes and colours close to those of the selection limits and can therefore serve as a test of our selection criteria (see also Sect.~\ref{select:cmd}). Field dwarf optical standards were also observed (LHS~234, vB~8, vB~10, LHS~2065, Kelu-1), though the final spectral classification method we adopted uses young optical standards instead (see Sect.~\ref{spec}). The observations were done with the long slit spectroscopy mode, with a slit width of 1$\arcsec$ or 2$\arcsec$ to better match the seeing conditions, resulting in a resolution of $\sim$500 and $\sim$300, respectively, across the spectral range. The individual exposure times were chosen according to the target's brightness and night conditions, and repeated with an ABBA pattern for posterior sky-subtraction. Standard A0 stars were observed at regular intervals and chosen to have an airmass matching that of the target within 0.1. The slit position was aligned with the parallactic angle. 

\begin{figure} 
   \centering
 \includegraphics[width=\columnwidth]{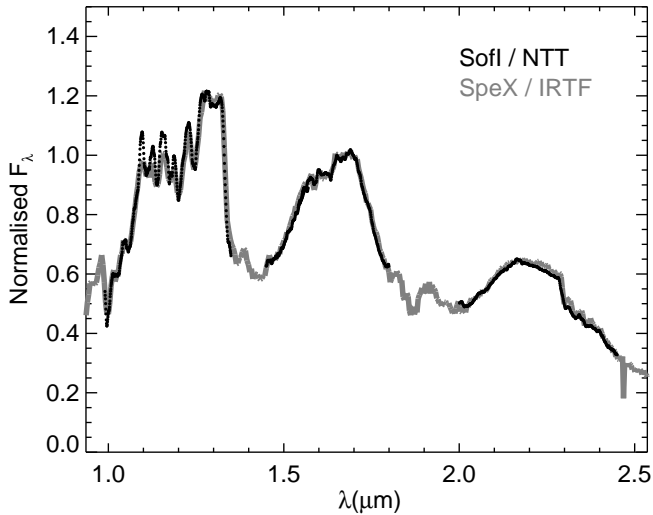}
   \caption{Spectra of Kelu-1 (an L dwarf optical standard) taken during the observing run with SofI / NTT (\emph{black}) and with the SpeX / IRTF (\emph{grey}) \protect{\citep[SpeX Prism Spectral Libraries]{Burgasser2007}}.}
   \label{kelu}
    \end{figure}

The data reduction was done first with the SofI pipeline developed and maintained by the Pipeline Systems Department at the European Southern Observatory (ESO). The 2D spectra were flat fielded, aligned, and co-added. The extraction of each spectrum was done with the \emph{APALL} routine in IRAF. All spectra were wavelength calibrated with a neon lamp. The telluric corrections were done by dividing each spectrum by that of the standard A0 stars observed at similar airmass and interpolated at the target's airmass. Relative fluxes were recovered with a theoretical spectrum of an A0 star \citep[taken from the ESO webpage] {Pickles1998} smoothed to the corresponding resolution. To remove the strong intrinsic hydrogen absorption lines from the spectra of the A0 standard stars, a linear interpolation was made across the lines that are more predominant at this resolution (the Paschen~$\delta$ line at 1.00~$\mu$m, Paschen~$\alpha$ at 1.09~$\mu$m, Paschen~$\beta$ at 1.28~$\mu$m, and the Brackett series lines at 1.54, 1.56, 1.57, 1.59, 1.61, 1.64, 1.68, 1.74, 2.17~$\mu$m). The final spectra were not flux-calibrated, but simply normalised to their average flux in the 1.67-1.71~$\mu$m region. The excellent agreement in the overlapping region of the spectra taken with the blue and red grisms shows that no further calibrations are needed. Figure~\ref{kelu} shows the comparison of the SofI / NTT spectrum of Kelu-1 (an L dwarf optical standard) taken during this observing run with a spectrum of the same object taken with the SpeX spectrograph \citep{Rayner2003} mounted on the 3~m NASA Infrared Telescope Facility, provided in the SpeX Prism Spectral Libraries\footnote{Available at http://www.browndwarfs.org/spexprism/}  \citep{Burgasser2007}. There is a good agreement in the spectral features and the two spectra exhibit similar relative fluxes between the three photometric bands.

\subsection{The NICS / TNG  spectroscopic follow-up}
\label{data:tng}
A shorter observing run was conducted at the Telescopio Nacional Galileo (TNG, La Palma, Observatory Roque de Los Muchachos) with the Near Infrared Camera and Spectrograph (NICS), a low-resolution spectrograph working in the near-IR regime. The observations took place during two half nights on 17-18 May 2009, and three of our candidate members with magnitudes of  \emph{K$_{s}$}$\sim$13 were observed (in addition to the main observing programme, which did not directly concern this study), as well as the field dwarfs vB~10 and LHS~2924. We used the \emph{JK'} grism, with a wavelength range from 1.15 to 2.23~$\mu$m, resulting in a resolution power of $\sim$350. The observations were done with the 1$\arcsec$ slit aligned at the parallactic angle, and A0 standard stars were observed for telluric corrections. Standard IRAF routines were employed to reduce the data, in an analogous way to that described in the previous section. The spectra were normalised to their average flux in the region between 1.67-1.71~$\mu$m.

\section{Archival data}
\label{archdata}
In order to complement our selection criteria of young stellar objects in $\rho$~Oph and to better characterise the new candidate and confirmed members, we made extensive use of multi-wavelength data recovered from different archives. The datasets used and the criteria employed to extract reliable samples are briefly described in this section.

   \begin{figure*} 
   \centering
 \includegraphics[width=\linewidth]{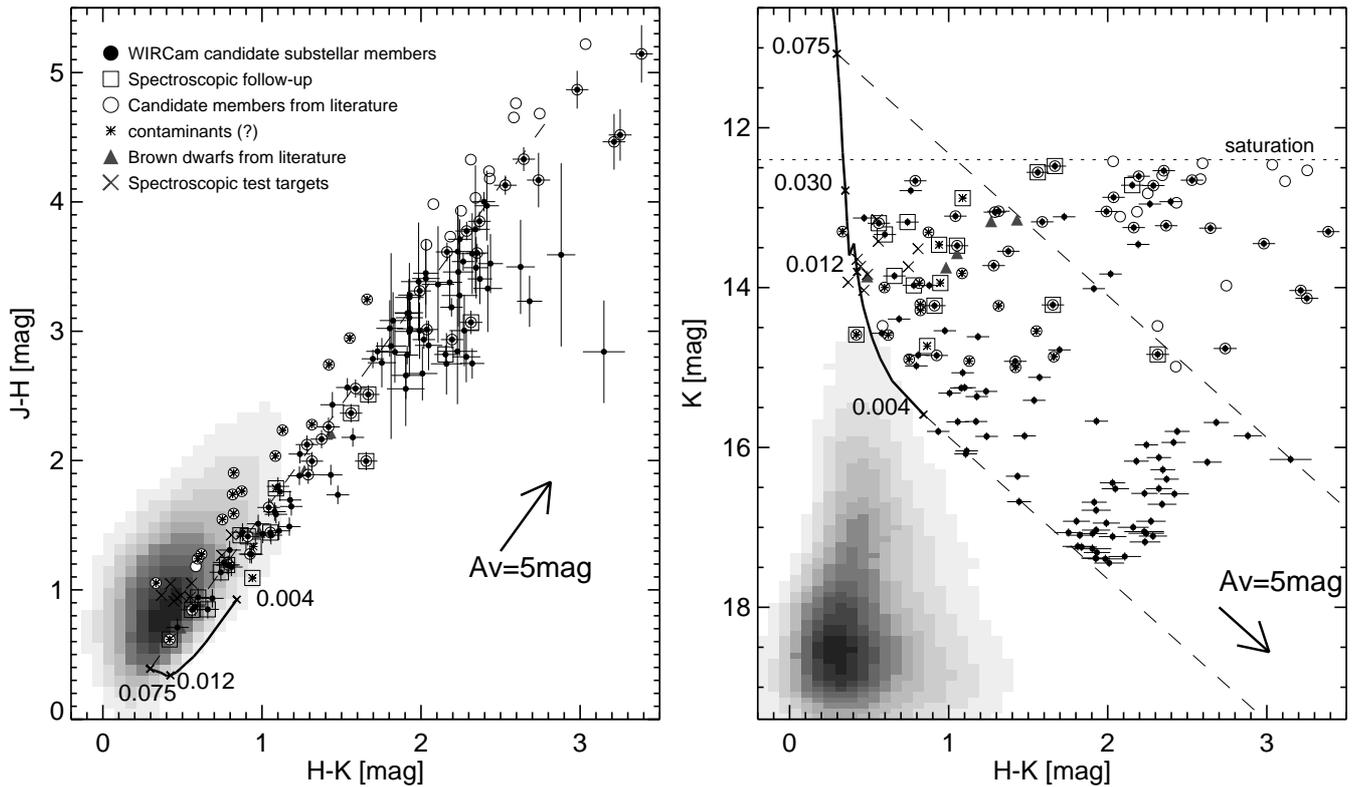}
   \caption{\emph{J}-\emph{H}~vs.~\emph{H}-\emph{K$_{s}$} colour-colour diagram (\emph{left}) and \emph{K$_{s}$}~vs~\emph{H}-\emph{K$_{s}$} colour magnitude diagram (\emph{right}). Plotted over the density map for the entire WIRCam catalogue for $\rho$~Oph are the candidate members selected from near-IR photometry in this study (black filled circles), all the candidate members of the cluster from the literature present in the WIRCam catalogues (white circles, asterisks signal potential contaminants) including the previously known brown dwarfs from the literature (grey triangles). Candidate members selected in this study for the spectroscopic follow-up (squares) and test sources observed spectroscopically (crosses) are also shown. In both diagrams, the solid line represents the DUSTY 1 Myr isochrone labelled with solar masses (M$_{\sun}$) \protect{\citep{Chabrier2000}}, and the dashed lines the $\sim$75 and 4~M$_{\emph{Jup}}$ limits, with increasing amount of visual extinction.}
 \label{cmdwircam}
    \end{figure*}

\subsection{Spitzer Space Telescope: C2D survey}
\label{data:spitzer}
To complement this study, \emph{Spitzer} data from the C2D legacy project \citep[\emph{From Cores to Disks}]{Evans2003} were included. The $\rho$~Ophiuchi molecular cloud has been mapped with \emph{Spitzer}'s Infrared Camera \citep[IRAC]{Fazio2004} in the 3.6, 4.5, 5.8 and 8.0~$\mu$m bands over a region of 8.0~deg$^{2}$ and with the Multiband Imaging Camera \citep[MIPS]{Rieke2004} in the 24 and 70~$\mu$m bands over a total of 14.0~deg$^{2}$ \citep{Padgett2008}, which encompass the WIRCam field in its totality. The data were retrieved from the C2D point-source catalogues of the final data delivery \citep{Evans2005}, using the NASA/ IPAC Infrared Science Archive\footnote{Available at http://irsa.ipac.caltech.edu/}. 

All fluxes were converted to magnitudes using the following zero points: 280.9$\pm$4.1, 179.7$\pm$2.6, 115.0$\pm$1.7, 64.1$\pm$0.94~(Jy), for the 3.6, 4.5, 5.8 and 8.0~$\mu$m IRAC bands, respectively, and 7.17$\pm$0.11~(Jy) for the 24~$\mu$m MIPS band. Only sources with magnitude errors below 0.3 magnitudes as well as detections above 2~$\sigma$ were kept. The \emph{Spitzer} catalogues were merged with the WIRCam detections catalogue, requiring the closest match to be within 1$\arcsec$. A counterpart was found for $\sim$15,000 objects that were detected in one or more mid-IR bands. Infrared excess around young stellar objects (YSOs) is a direct evidence of discs and their detection is commonly used as a youth indicator \citep{Haisch2001}. These data are relevant to assess the likelihood of membership for the candidate members as well as to characterise their morphological properties (Sect.~\ref{comp:spitzer}). 

\subsection{Subaru Telescope: $i'$ and $z'$ band archival data}
\label{data:subaru}
We searched the \emph{Subaru} Mitaka Okayama Kiso Archive system \citep[SMOKA][]{Baba2002} for \emph{Subaru}  Prime Focus Camera  \citep[Suprime-Cam]{Miyazaki2002} optical images overlapping with the WIRCam/CFHT survey. We found one overlapping field observed in the Sloan $i'$-band on 20 June 2007, and two fields observed in the Sloan $z'$-band on 16-17 April 2004. Table~\ref{table:2} gives a summary of the observations. Each field was observed in dithering mode to effectively compute and remove the sky. Weather conditions on Mauna Kea on 20 June 2007 and 17 April 2004 were photometric, as reported by the CFHT Skyprobe atmospheric attenuation measurements \citep{Cuillandre2004}. No data are available for 16 April 2004, but the quality and depth of the images suggests that the weather was similarly good. Seeing as measured in the images ranged from 0$\farcs$6\ to 0$\farcs$8 in the $i'$ and $z'$-band images. 

The 10 individual CCD of the Suprime-Cam mosaic were processed with the standard reduction procedure with the recommended SDFRED package \citep{Yagi2002,Ouchi2004}. The programme SDFRED performs overscan and bias subtraction, flatfielding, distortion correction, atmospheric dispersion correction, sky subtraction, masking vignetted regions, and alignment and co-addition. A sixth order astrometric solution was computed using 2MASS counterparts. The final accuracy is expected to be better than 0$\farcs$2. The programme SExtractor was used to identify all sources brigther than the 3-$\sigma$ local standard deviation over at least 3 pixels. The absolute zeropoint for each CCD was derived from the observation of a SDSS secondary standard field \citep[SA~110,][]{Schmidt2002} observed the same night. They are given in Table~\ref{table:2} and agree with the \citet{Miyazaki2002} measurements within 0.12~mag and 0.02~mag in the $i'$ and $z'$-band, respectively. The chip-to-chip offsets were computed from the median flux of domeflat images.

   \begin{table*}
    \begin{minipage}[t]{\linewidth}
    \centering             
\caption{Journal of the Suprime-cam / \emph{Subaru} observations.} 
\begin{tabular}{lcccccc}\hline\hline
Pointing &  RA               & Dec.              & Date                &   Filters         & Exp. Time                & Zeropoint \\
               &  (J2000)        & (J2000)         &                          &                      &                                   & [ABmag] \\
\hline
1              & 16 27 00      & -24 38 21     & 20 June 2007  & Sloan $i'$   &  20$\times$80~s     &  28.04$\pm$0.03 \\
2              & 16 27 06      & -24 13 30     & 17 April 2004  & Sloan $z'$   & 16$\times$240~s   &  27.03$\pm$0.03 \\
3              & 16 25 06      & -24 38 30     & 16 April 2004  & Sloan $z'$   &  21$\times$200~s  &  27.04$\pm$0.04 \\
    \hline                                 
   \label{table:2} 
   \end{tabular}
   \end{minipage}
   \end{table*}


\section{Selection of substellar candidate members of $\rho$~Oph}
\label{select:cmd}

\subsection{Colour-colour and colour-magnitude diagrams}

The primary criteria used to select candidate members in the substellar regime is to compare the positions of all the WIRCam sources in various colour-colour and colour-magnitude diagrams with the predictions from the models of the YSOs colours. In the first iteration and in an attempt not to exclude any possibly interesting candidates, we performed no filtering on the initial catalogue of elongated objects (galaxies, nebulosities) or objects that could have their photometry affected by instrument artifacts. This step was done later with more stringent criteria to ensure the quality of the selected candidates. Evolutionary models are known to become increasingly uncertain at younger ages and lower masses, which is the case for our survey, and a more accurate way to select candidate members based on photometry diagrams is to use observed colours of known YSOs instead. \citet{Luhman2010} empirically determined intrinsic colours  for young stars and brown dwarfs, which the authors present in the 2MASS photometric system. Given the differences between 2MASS and the WIRCam filter system, which cause large differences in colour, and because to date no colour transformation equations between the two systems have been derived, we cannot use this approach though. We rely on evolutionary models for the selection of candidate substellar members, noting that when comparing the observed colours of YSOs from \citet{Luhman2010} to the Dusty model (2MASS filters, 1Myr), we find the differences to be of the order of our photometric colour errors. Furthermore, using the models we recover the previously known brown dwarfs and the majority of candidate members from the literature present in the WIRCam catalogues (see Sect.~\ref{comp:surveys} for a detailed discussion). 

In the first step, candidate members of $\rho$~Oph were selected if their colours fell redward from the model isochrones in the \emph{J} vs. \emph{J}-\emph{H},  \emph{J} vs. \emph{J}-\emph{K$_{s}$}, and \emph{K$_{s}$} vs. \emph{H}-\emph{K$_{s}$} colour-magnitude diagrams. From these, only objects that had colours consistent with them being young and substellar in the \emph{J}-\emph{H}~vs.~\emph{H}-\emph{K$_{s}$} colour-colour diagram (Fig.~\ref{cmdwircam}) were kept. The 1~Myr Dusty isochrone \citep{Chabrier2000} computed for WIRCam/CFHT \emph{J}, \emph{H}, and \emph{K$_{s}$} filters was used down to a temperature of $\sim$1700~K, shifted to a distance of 130~pc. The adopted distance to $\rho$~Oph is a median value to that of several distance estimates existing in the literature, which indicates that distances to different regions of the cloud can vary from 120 to 145~pc  \citep{Lombardi2008,Mamajek2008}. According to the models, the substellar limit is at \emph{J}$\sim$11.8, \emph{K}$\sim$11.1, and \emph{J}-\emph{H}$\sim$0.3. This selection holds 178 substellar candidates. From this list, sources were removed which had a flux-radius value (flux-radius is a SExtractor parameter that measures the radii of the PSF profile at which the flux is 50\% from its maximum) inconsistent with stellar profiles, i.e. close to zero or much larger than the average PSF FWHM measured on the images. This ensured that instrumental artifacts (bad pixels, cosmics) and elongated sources (galaxies, nebulosities), respectively, were correctly discarded. The remaining sources have been visually inspected, and a further rejection criterion was implemented to exclude detections susceptible of having bad photometry, like those at the edge of the field, in the overlapping regions between detectors (mostly in the detectors edges), or in zones of bright reflection nebulae. 

We removed from the candidate list five young brown dwarfs (GY~11, 64, 141, 202, CRBR~31, see also Table~\ref{bd}), which have already been spectroscopically confirmed as members \citep{Luhman1997,Wilking1999,Cushing2000,Natta2002}. There are several objects which have previously been associated with the cloud but lack a spectroscopic confirmation (see also Sect.~\ref{comp:surveys}), in particular from previous IR surveys from \citet{Greene1992}, \citet{Strom1995}, and \citet{Bontemps2001}; members identified through X-ray emission \citep{Imanishi2001,Gagne2004}; and candidate members proposed based on their \emph{Spitzer} colours \citep[see Sect.~\ref{comp:spitzer}]{Padgett2008,Wilking2008,Gutermuth2009}. These were kept in our catalogue and are referenced accordingly when mentioned. We also removed objects from the list of candidate members that turned out to be field star contaminants from our spectroscopy follow-up (see Sect.~\ref{comspec}, the coordinates and near-IR magnitudes for these sources are given in Table~\ref{cont}), as well as three sources observed by \citet{Marsh2009}, which were found not to be substellar (identifiers in that study are $\#$1307, $\#$2438, and $\#$2403). The final list contains 110 substellar candidates selected from near-IR photometry alone, which are listed in Table~\ref{stars}.

\begin{table}
\caption{WIRCam data for field star contaminants.}           
\begin{tabular}{c c c c c}        
\hline            
\hline
RA& Dec & \emph{J} & \emph{H} & \emph{$K_{s}$} \\
~ &   ~      & (mag)        & (mag)          & (mag)        \\
\hline
16 25 15.92 &	-24 25 10.8 & 15.50$\pm$0.05 & 14.40$\pm$0.05 & 13.46$\pm$0.05  \\
16 25 49.93 &	-24 13 43.7 & 17.02$\pm$0.05 & 15.60$\pm$0.05 & 14.73$\pm$0.05 \\
16 26 22.47 &	-24 37 24.2 & 16.22$\pm$0.05 & 14.89$\pm$0.05 & 13.94$\pm$0.05  \\
16 26 46.12 &	-24 21 53.8 & 15.75$\pm$0.05 & 13.97$\pm$0.05 &  12.88$\pm$0.05 \\
\hline
 \label{cont} 
\end{tabular}
\end{table}

During the spectroscopic follow-up, we observed nine sources that were not part of the substellar candidates list. These sources were chosen to pass the selection criteria in the various CMDs, but to have positions in the colour-colour magnitude diagram near to the reddening line extended from the 75~M$_{\emph{Jup}}$ model colour, which we used for the selection of candidates. The sources have magnitudes less than $\sim$14 in the \emph{K$_{s}$} band. All these turned out to be field dwarfs, further supporting the limits used in our selection criteria, in particular at this magnitude range. These sources are also shown in Fig.~\ref{cmdwircam} (\emph{crosses}).

\addtocounter{table}{1} 

\subsection{Comparison to previous surveys of the $\rho$~Ophiuchi molecular cloud}
\label{comp:surveys}
To better evaluate the quality of our images, detection methods, and the consistency of our candidate selection criteria, we compared our results to those of previous surveys of $\rho$~Oph. In a recent compilation of previous studies in the literature, \citet{Wilking2008} gathered a list of 316 confirmed or candidate members of the cluster, from which 295 have positions on sky within the field of our survey. The majority of these are part of the initially extracted WIRCam catalogues, with only 13 objects from the literature missed by the detection algorithm. From these, nine sources (ISO-29, 31, 60, 85, 90, 99, 125, 137, 144 \citep{Bontemps2001}; [GY92]-167, 168 \citep{Greene1992}; HD 147889; CRBR-36 \citep{Comeron1993}) are either extremely saturated in the WIRCam images or in the vicinity of those and other bright reflection nebulae, and therefore impossible to detect in the saturated pixels. The remaining four (ISO-60, 85, 90, 99) were previously associated with the cloud either from their X-ray emission or mid-IR excess. However, there is no signal detected in the \emph{J}-band of the WIRCam images, hence they are not present in the combined \emph{JHK$_{s}$} catalogue. Finally, sixty-five of the previously known or candidate YSOs are within the magnitude range of our near-IR candidate selection, and 35 of those fall in the substellar region (including the five brown dwarfs mentioned in the previous section), further supporting our selection criteria. Objects detected by our extraction algorithm but not within the magnitude range adopted for our candidate selection include seven sources that are too faint in the \emph{J} band and have no reliable magnitude measurement, and 211 objects that are brighter than the survey saturation limit for one or more near-IR bands. 

While comparing the positions of the candidate members from the literature from the compiled list of \citet{Wilking2008} in the near-IR diagrams, we found inconsistencies between the colours of fifteen sources and the expected colours of YSOs. In the colour-magnitude diagrams, these sources are in the substellar regime region (we caution that this is only an approximation, given the known errors in the models). However, in the colour-colour diagram, they show colours consistent with those of stars. One of these sources is an edge-on disc \citep[known as the \emph{Flying Saucer}]{Grosso2003}. Given the extended profile of this source, it is likely that its PSF photometry done on the WIRCam images has larger errors, because the parameters were optimised for point-source photometry. The remaining 14 sources include ROXC J162821.8-245535, which according to \citet{Wilking2008} has been assigned membership based on X-ray emission (unpublished), and 13 sources classified as candidates from the analysis of IRAC data done by \citet{Wilking2008}, where several diagnostics for the detection of mid-IR excess were used (identification numbers used in that work for these sources are IRAC~20, 746, 763, 830, 831, 869, 901, 1016, 1086, 1212, 1343, 1350, 1401). From these, only one source (GY~376, or IRAC~746) is present in the list of candidate members of $\rho$~Oph from \citet{Gutermuth2009} who used the same \emph{Spitzer} dataset, and none have been otherwise previously associated with the cloud. Since \citet{Wilking2008} have not provided details of their reduction of the data, or mid-IR magnitudes for the new candidates we cannot further comment on the validity of their selection. However, according to our near-IR dataset, these sources seem inconsistent with being members of $\rho$~ Oph (see also Sect.~\ref{comp:spitzer}).

Additionally, we included in our study recent results not found in the compilation from \citet{Wilking2008} from the DROXO survey \citep[Deep Rho Ophiuchi XMM-Newton Observation]{Sciortino2006}, which consists of a very deep exposure (total exposure time is 515~ksec) taken with the European Photon Imaging Camera \citep[EPIC]{Struder2001,Turner2001} on board the XMM-\emph{Newton} satellite \citep{Jansen2001}, and covering a region of $\sim$0.2~deg$^{2}$ of the $\rho$~Ophiuchi cluster. The data reduction and main results from this survey can be found in \citet{Giardino2007}, \citet{Flaccomio2009}, and Pillitteri et al. (2010, \emph{in prep.}). A total of 111 X-ray emitting sources are reported in their studies. The sensitivity of the DROXO survey and area covered are much lower than that of our survey. We found a positional match within 2~$\sigma$ from the source positional errors for two of our candidate members.

We also compared the WIRCam data with the \emph{K$_{s}$} photometry presented by \citet{Marsh2009} for the seven candidate YSOs in $\rho$~Oph observed spectroscopically in that study (Table~\ref{marsh}). \citet{Marsh2009} derive near-IR photometry from stacking deep integration \emph{J}, \emph{H}, and \emph{K$_{s}$} images from the 2MASS calibration scans. We found a good agreement between their 2MASS measurements and the WIRCam photometry for four sources ($\#$1449, $\#$1307, $\#$2438, $\#$2403) with magnitude differences between 0.02 and 0.23 magnitudes. But for the remaining three sources we found larger variations, with differences in \emph{K$_{s}$} magnitude of 0.4, 1.42, and 1.57 for sources $\#$2974, $\#$4450, $\#$3117. Two of these sources ($\#$2974, $\#$3117), were also detected in the WFCAM/UKIRT images from \citet{Alvesdeoliveira2008}, and their magnitudes have a difference of 0.14 and 0.17 to those of the WIRCam, which is of the order of their photometric errors at this magnitude range and seems to indicate that our measurements are correct. Of particular interest is the result for source $\#$4450, classified by \citet{Marsh2009} as a young T2 dwarf, which may be the youngest and least massive T dwarf  observed spectroscopically so far. From our images we derived a \emph{K$_{s}$} magnitude of 19.14$\pm$0.20, whereas \citet{Marsh2009} reported \emph{K$_{s}$}=17.71. Assuming the parameters derived in the spectral analysis of \citet{Marsh2009} are correct and repeating the same calculation as the authors to estimate the distance, we arrive at a $\pm$1$\sigma$ range in distance of 137 to 217~pc using the WIRCam magnitude value. This seems to suggest that this object is behind the $\rho$~Ophiuchi cloud and could instead be part of the Upper Sco association \citep{deGeus1989} located at $\sim$145~pc \citep{deBruijne1997} and an estimated age of 5~Myr \citep{Preibisch1999}. As claimed by \citet{Marsh2009}, in that case the estimated mass for this brown dwarf would still be $\le$~3 Jupiter masses, according to the models. We could not derive a magnitude for the WIRCam \emph{H} band because the object's position is coincident with an artifact caused by the guiding star. In the \emph{J} band, we derived a magnitude of 21.32$\pm$0.35. Its \emph{J}-\emph{K$_{s}$} colour is consistent with a T dwarf reddened by the extinction measured spectroscopically by \citet{Marsh2009}.

   \begin{table*}
    \begin{minipage}[t]{\linewidth}
    \centering             
\caption{WIRCam data for sources in \protect{\citet{Marsh2009}}.}           
\begin{tabular}{c c c c c c}        
\hline            
\hline
ID (as in \citet{Marsh2009}) & RA& Dec & \emph{J} & \emph{H} & \emph{$K_{s}$} \\
~ &   ~      & ~      & (mag)        & (mag)          & (mag)        \\
\hline
2438  & 16 27 09.37  & -24 32 14.9  & 22.21$\pm$0.39  & 19.53$\pm$0.14  & 16.97$\pm$0.05      \\
2974  & 16 27 16.74  & -24 25 39.0  &                               &                                 &  17.26$\pm$0.06      \\
3117  & 16 27 17.68  & -24 25 53.5  &                                &  18.64$\pm$0.07 & 17.27$\pm$0.05       \\
2403  & 16 27 21.63  & -24 32 19.2  & 20.86$\pm$0.10   & 18.17$\pm$0.06  & 16.50$\pm$0.05       \\
4450  & 16 27 25.35  & -24 25 37.5  & 21.32$\pm$0.35   &                                & 19.14$\pm$0.20       \\
1449  & 16 27 30.36  & -24 20 52.2  & 19.59$\pm$0.06   &  16.96$\pm$0.05 & 15.65$\pm$0.05        \\
1307  & 16 27 32.89  & -24 28 11.4  &  21.41$\pm$0.15   &  17.44$\pm$0.05  & 14.97$\pm$0.05       \\
\hline
\label{marsh}
   \end{tabular}
   \end{minipage}
   \end{table*}


\section{Photometric properties of the candidate members}
\label{discussion_phot} 

\begin{figure}
   \centering
 \includegraphics[width=\linewidth]{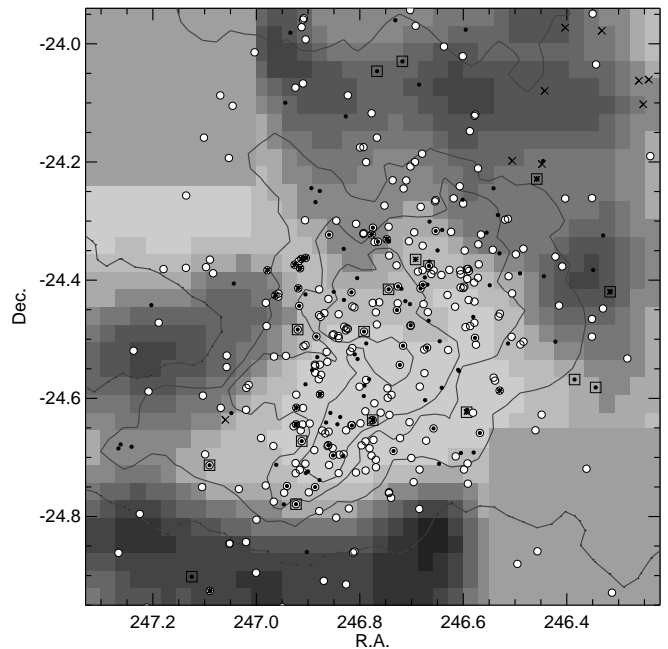}
   \caption{Spatial distribution of the substellar candidate members of $\rho$~Ophiuchi in this study and previously known candidate and confirmed members of the cloud as compiled by \protect{\citet{Wilking2008}}, superposed on the density map of all WIRCam detections with contours from the NICER extinction map. Symbols are the same as in Fig.~\ref{cmdwircam}.}
   \label{image}
    \end{figure}

We used multiwavelength data to characterise the candidate members presented in Table~\ref{stars}. For each candidate, a flag was included to indicate additional information: previously suggested candidate member in the literature (Sect.~\ref{comp:surveys}), candidate selection confirmed from optical counterpart (Sect.~\ref{comp:optical}), mid-IR excess as determined from \emph{Spitzer} diagrams (Sect.~\ref{comp:spitzer}), variability behaviour (Sect.~\ref{comp:var}), or their membership confirmed spectroscopically from this study (Sect.~\ref{spec}). Figure~\ref{image} shows the position on sky of the candidate members and the known members of $\rho$~Oph, superposed on the density map of all WIRCam detections and the contours from the extinction map of the $\rho$~Ophiuchi cloud provided by the COMPLETE\footnote{Available at http://www.cfa.harvard.edu/COMPLETE}  project \citep{Ridge2006,Lombardi2008}, and computed with the NICER algorithm \citep{Lombardi2001} using 2MASS photometry.

   \begin{figure*}
   \centering
 \includegraphics[width=\linewidth]{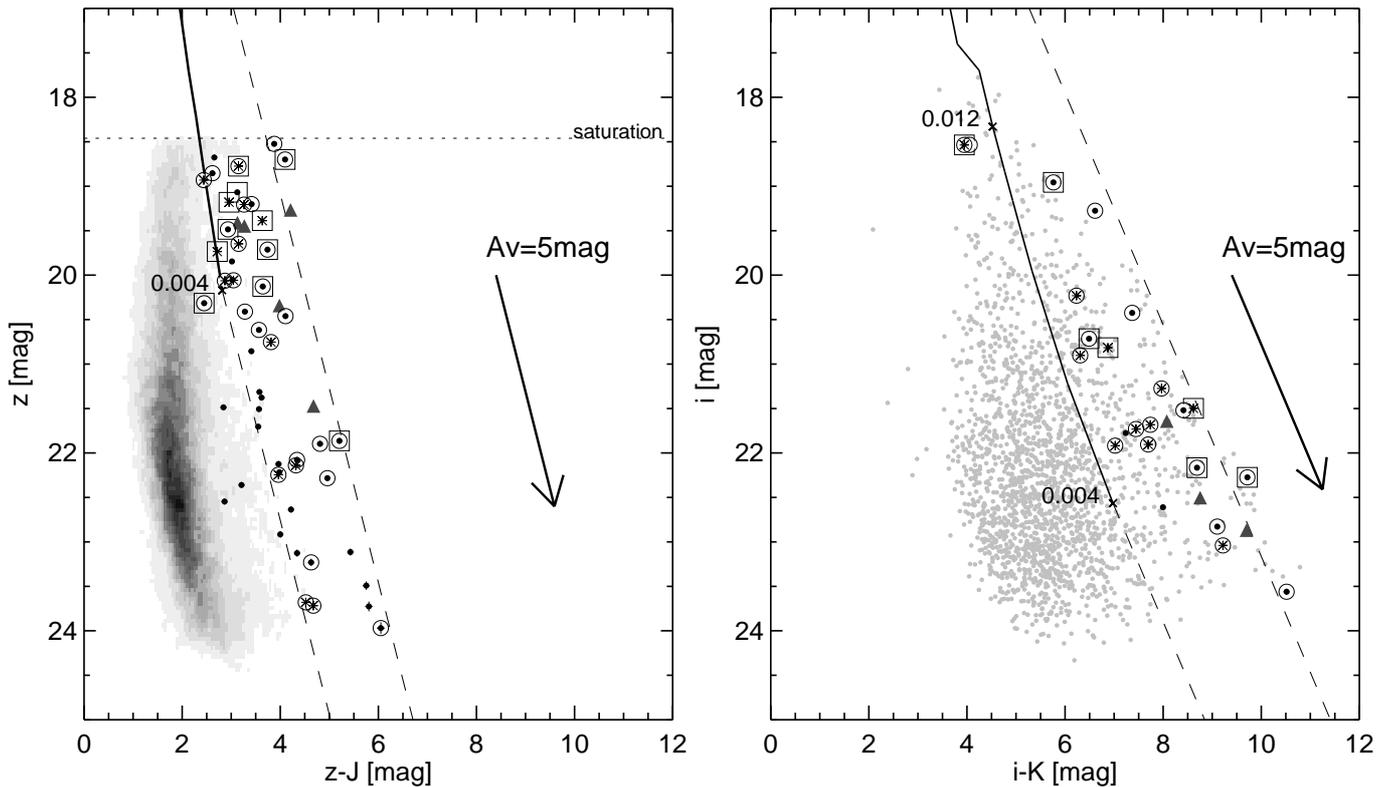}
   \caption{\emph{$z'$}~vs~\emph{$z'$}-\emph{J} (\emph{left}) and \emph{$i'$}~vs~\emph{$i'$}-\emph{J} (\emph{right}) colour magnitude diagrams. In both diagrams the solid line represents the DUSTY 1 Myr isochrone labelled with solar masses (M$_{\sun}$) \protect{\citep{Chabrier2000}}, and the dashed lines the $\sim$75 and 4~M$_{\emph{Jup}}$ limits, with increasing amount of visual extinction. Symbols are the same as in Fig.~\ref{cmdwircam}.}
   \label{subaru}
    \end{figure*}

\subsection{Near-IR and optical CMDs }
\label{comp:optical}

For the candidate members with a counterpart in the $i'$ and $z'$-bands from the \emph{Subaru} telescope, we could further test our near-IR selection criteria. A visual inspection of the match between candidate members and the optical counterparts was performed to ensure the quality of the positional association. The candidate members without an optical counterpart were either to faint in the the $i'$ or/and $z'$-bands or were not in the area covered by the optical images. Figure~\ref{subaru} shows the colour-magnitude diagram combining infrared and optical data together with the theoretical isochrone from the DUSTY models \citep{Chabrier2000} for 1~Myr age, shifted to 130~pc. All but five candidates with either $i'$ and/or $z'$-band photometry show colours consistent with those predicted by evolutionary models within the photometric measurement errors, and more important with those of the previously known brown dwarfs in $\rho$~Oph. Our selection criteria are therefore confirmed for the majority of the sources present in these diagrams. Furthermore, GY~201 shows colours that are too blue when compared to the isochrones or to the other candidates and members.

   \begin{figure*}
   \centering
 \includegraphics[width=\linewidth]{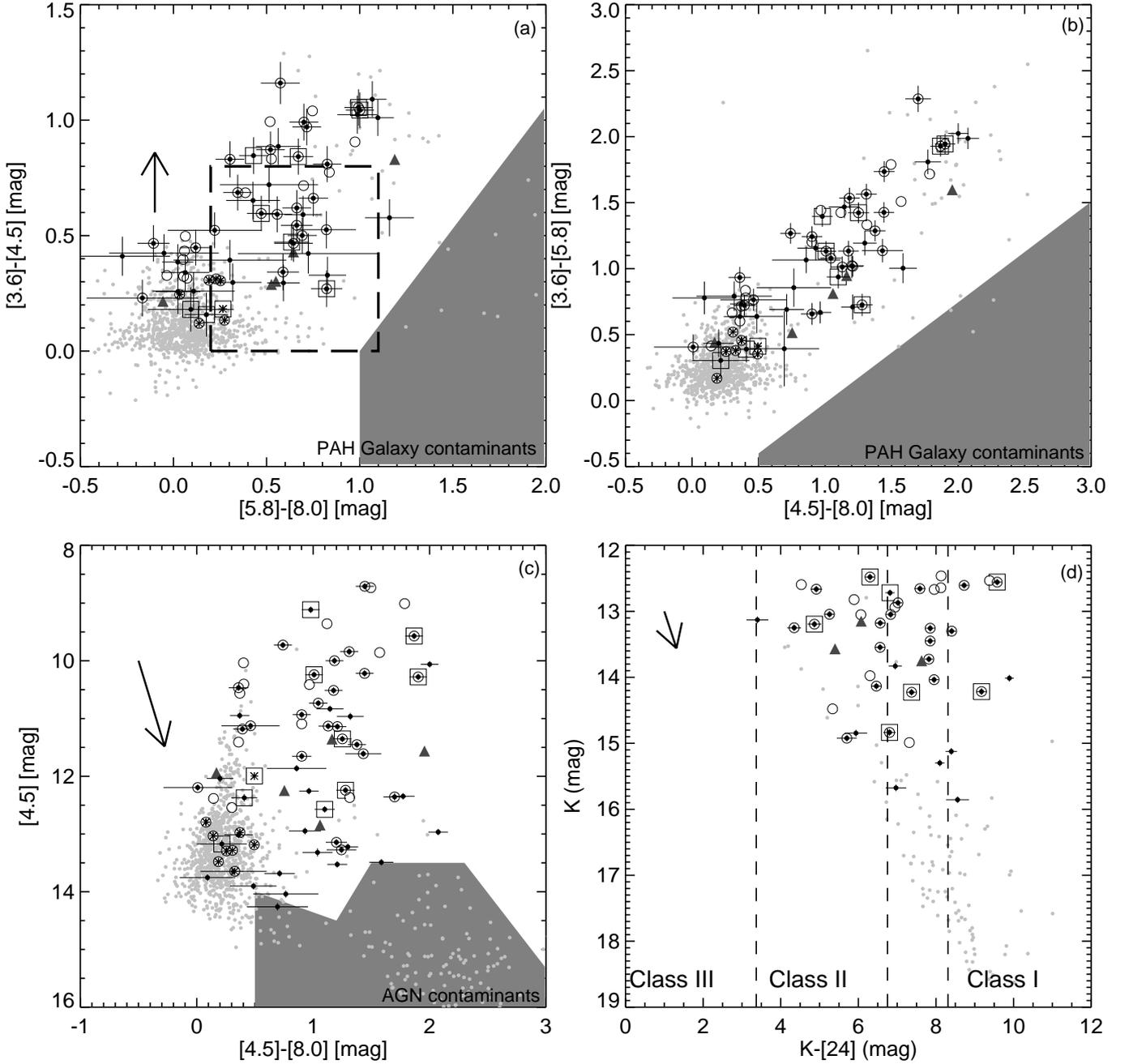}
   \caption{IRAC / \emph{Spitzer} colour-colour diagrams (panels (a), (b), and (c)), and WIRCam/CFHT and MIPS/\emph{Spitzer} colour-magnitude diagram. The various diagrams are used to separate young stars of different SED classes, and identify possible extragalactic contaminants. Symbols are the same as in Fig.~\ref{cmdwircam}.}
   \label{spitzer}
    \end{figure*}

\subsection{Mid-IR excess from \emph{Spitzer} data}
\label{comp:spitzer}
Young stellar objects can show infrared emission, which originates from dusty envelopes and circumstellar discs surrounding the central object. Mid-IR data from the IRAC and MIPS \emph{Spitzer} cameras allow these objects to be studied at wavelengths where the excess contribution from discs and envelopes is predominant. With several colour-colour and colour-magnitude diagrams (Fig.~\ref{spitzer}) we could further characterise our list of candidate members. Previous work on Spitzer mid-IR observations of the $\rho$~Ophiuchi cluster has been published by \citet{Padgett2008} and \citet{Gutermuth2009}, where hundreds of candidate members are uncovered over a much larger area than the WIRCam/CFHT survey. 

The IRAC colour-colour diagram ([3.6]$-$[4.5] vs. [5.8]$-$[8.0]) in the panel (a) of Fig.~\ref{spitzer} can be used as a tool to separate young stars of different classes \citep{Allen2004,Megeath2004} and to reject sources consistent with galaxies dominated by PAH emission and narrow-line AGN \citep{Gutermuth2009}. Centred in the origin are sources which have colours consistent with stellar photospheres and have no intrinsic IR-excess. These can be foreground and background stars, but also Class~III stars with no significant circumstellar dust. In this region of the colour-colour plane it is impossible to differentiate between young stars and contaminants. Another preferred region for objects in the diagram is located within the box defined by \citet{Allen2004}, which represents the colours expected from models of discs around young, low-mass stars.  Finally, from models of infalling envelopes,  \citet{Allen2004} predict the colours of Class~I sources to have ([3.6]$-$[4.5])~$>$~0.8 and$/$or ([5.8]$-$[8.0])~$>$~1.1. Thirty seven of our candidate members have good photometry in the four IRAC bands and are displayed in the diagram. The remaining candidates are either too faint in the IRAC images or have detections in one of the four bands that did not match the quality criteria we applied (see Sect.~\ref{data:spitzer}). From those, 21 were previously associated with $\rho$~Oph. All the candidates in panel (a) have colours consistent with Class~II or Class~I young objects. Furthermore, none of the candidates falls in the preferred region defined by the colours of possible extragalactic contaminants (see Sect.~\ref{contamination}), which is also confirmed from the diagram (b). Panel (c) can be used to estimate contamination levels by broad-line AGN and is further discussed in Sect.~\ref{contamination}.

Twenty six of the candidate members have a detection at 24~$\mu$m and are displayed in panel (d).  Following the \citet{Greene1994} mid-IR classiÞcation scheme (based on the $\alpha$ index), YSOs will lay on defined areas of the diagram, namely \emph{K$_{s}$}$-$[24] $>$ 8.31 for Class I, 6.75 $<$ \emph{K$_{s}$}$-$[24] $<$ 8.31 for flat-spectrum objects, 3.37 $<$ \emph{K$_{s}$}$-$[24]  $<$ 6.75 for Class II, and \emph{K$_{s}$}$-$[24]  $<$ 3.37 for photospheric colours. All our candidates with a detection at 24~$\mu$m are consistent with being young according to this classification scheme. 

\subsection{Near-IR variability of YSOs}
\label{comp:var}
Variability is a characteristic of YSOs, and near-IR variability surveys in particular can probe stellar and circumstellar environments and provide information about the dynamics of the on going magnetic and accretion processes. With the Wide Field near-IR camera (WFCAM) at the UKIRT telescope, \citet{Alvesdeoliveira2008} conducted a multi-epoch, very deep near-IR survey of $\rho$~Oph to study photometric variability. They found 137 variable objects with timescales of variation from days to years and amplitude magnitude changes from a few tenths to $\sim$3 magnitudes. We found that 17 of our candidates show photometric variability (14 are included in the list of members compiled by \citet{Wilking2008}), which further supports their membership. From their list of candidate members found through photometric variability, 18 are outside our surveyed area, and 58 are saturated in the WIRCam images. From the variables that have magnitudes and colours consistent with our selection criteria for substellar candidate members of $\rho$~Oph, we recovered all but one. The source \object{AOC~J162733.75-242234.9} has large photometric variations in the near-IR (0.6 and 0.4 magnitudes in the \emph{H} and {K} band, respectively) and is detected in all the WIRCam images. But its position overlaps with an artefact in the \emph{J} band image, which may affect its photometry and explain the \emph{J}-\emph{H} colour found, which is too blue in comparison to the theoretical models. From our list of candidates, 93 do not appear in the list of near-IR variables. These include 12 candidates that are out of the field surveyed by \citet{Alvesdeoliveira2008} and 24 that are fainter than their completeness limits ($\sim$19 and 18 magnitudes, in \emph{H} and {K}, respectively). We therefore conclude that the remaining 57 candidate members (from which 13 have been previously associated with the cloud) did not show photometric variability in the near-IR on timescales from days to one year during the epochs surveyed in that study. 

\subsection{Contamination}
\label{contamination}

Photometrically selected samples of candidate members of young star-forming regions are prone to be contaminated by other objects with colours similar to those of YSOs. Only a spectroscopic analysis can reveal the true membership of the candidate members, but this requires large amounts of observing time in telescopes and that is not always achievable. The possible sources of contamination are extragalactic objects (like AGN or PAH galaxies), foreground and background field M dwarfs, and background red giants. We tried to estimate the level of contamination in our list of candidate members, calling to attention however that the membership of an individual source is ultimately dependent on its position in the field, because for a large part of the area of sky surveyed, the cloud's extinction will substantially shield any background contamination. We removed from the discussion the 30 candidates that were previously associated with the cloud because they gather an ensemble of properties from different surveys that further supports their membership (see Sect.~\ref{comp:surveys}), as well as the spectroscopically confirmed members in this study, leaving 70 candidates.

\subsubsection{Extragalactic contaminants}

In an attempt to characterize the extragalactic contamination levels in YSO samples selected on the basis on IR-excess emission, in particular using \emph{Spitzer} observations, \citet{Gutermuth2009} explored the same data as \citet{Stern2005} to select active galaxies and compare them to YSO selection methods. In particular, \citet{Stern2005} found that PAH-emitting galaxies have colours that are confined to specific areas in most of the IRAC colour-colour diagrams. These regions were adopted by \citet{Gutermuth2009} to filter out these contaminants, and they are depicted in panels (a) and (b) of Fig.~\ref{spitzer}. None of the candidate members present in this diagram falls into either of these regions, indicating that these group of candidates is most likely not affected by contamination from star-forming galaxies.

Another source of contamination are broad-line AGN which have mid-IR colours similar to those of YSOs \citep{Stern2005}. With the IRAC diagram panel (c) in Fig.~\ref{spitzer} we try to provide an estimate for contamination level from AGN, following the methodology from \citet{Guieu2009} for underredened colours \citep[based on][]{Gutermuth2009}. The region plotted in the diagram (\emph{in grey}) shows the area in the [4.5] vs. [4.5]$-$[8.0] colour space consistent with AGN-like sources. \citet{Gutermuth2009} found that while applying this cutoff significantly improved the extragalactic filtering of catalogues, some residual contamination is still expected. Three of the YSO candidates in these diagram fall in the contamination area and are signaled out as possible contaminants in Table~\ref{stars}. However, only 40 of the candidates have detections at 4.5 and 8~$\mu$m. If we extrapolate this to the candidate list, we estimate a contamination level of $\sim$5 extragalactic sources, a conservative upper-limit because we are not taking into account the cloud's extinction.

\subsubsection{Galactic contaminants}
In the Galaxy, giants are an important source of contamination. However, taking in account the high galactic latitude of the $\rho$~Oph field (+16.7) and that background contamination is reduced due to cloud extinction, fewer giants are expected because we are surveying a region of the sky above the plane and bulge, which becomes dominated by the faint end of the dwarf luminosity function. 

We used the Besan\c{c}on model of galactic population synthesis \citep{Robin2003} to estimate the level of contamination by foreground late type objects and background red giants or extincted galactic sources. We retrieved a synthetic catalogue of sources within 0.78$^{2}$ toward the direction of our survey for distances in the range 0--50~kpc. Objects further away than 100~pc (hereafter background objects) were placed randomly within the field of our survey, and the corresponding extinction as given in the COMPLETE map was applied. The luminosities of objects closer than 100~pc (hereafter foreground objects) were not changed. Our selection algorithm was then applied to the output synthetic catalogue. According to these simulations, only one unrelated galactic source could have passed our selection criterion, indicating that the contamination by galactic sources must be low. The contaminant is an extincted (A$_{\emph{V}}$=7.5~mag) 1~Gyr old M8V dwarf located at 140~pc. The strong extinction in the region covered by our survey indeed most likely blocks the light of the majority of background sources up to the limit of sensitivity of the survey.


\section{Spectroscopic follow-up of candidate YSOs}
\label{spec}

 \begin{figure*}
   \centering
 \includegraphics[width=\linewidth]{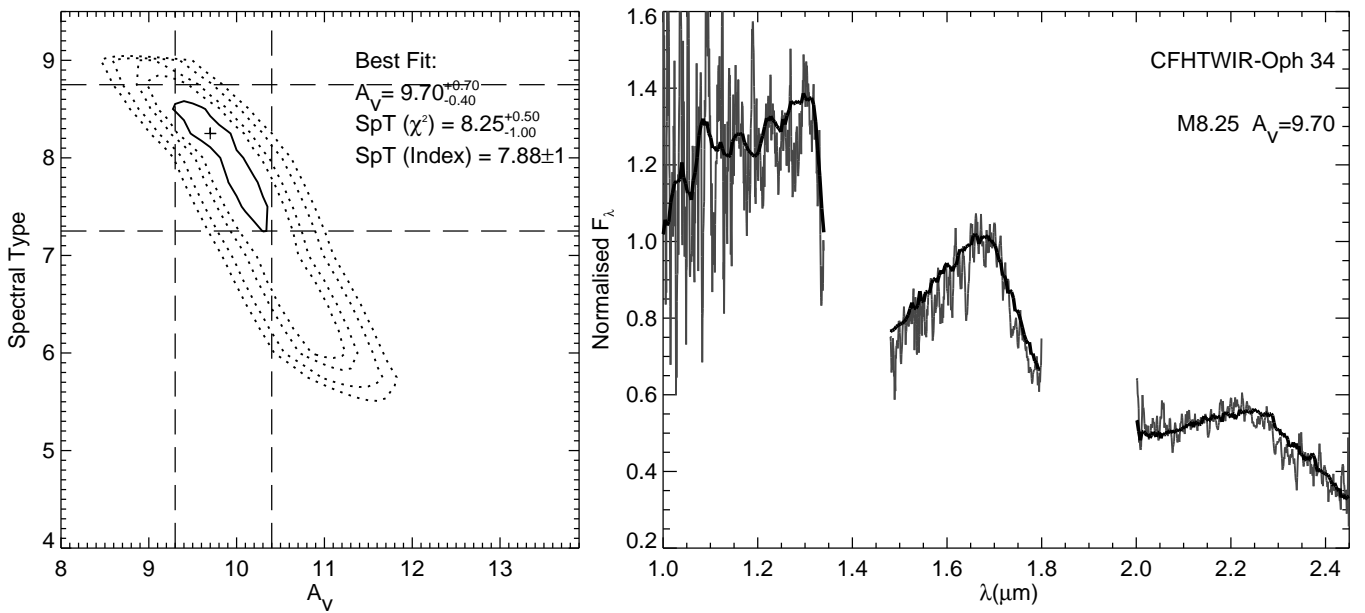}
   \caption{$\chi^{2}$ map obtained in the spectral type vs. A$_{\emph{V}}$ plane for one of the candidates (CFHTWIR-Oph 34), together with the best-fit solution. The spectrum shows a very good match to that of an intermediate spectrum between an M8 and M9 young brown dwarf of Taurus (1Myr). }
\label{chi2fit}
    \end{figure*}

We obtained near-IR spectra for 16 candidate members of $\rho$~Oph, chosen from our WIRCam/CFHT survey, and for GY~201 a candidate member from the literature. The low resolution and modest signal-to-noise of the spectra (S/N$\sim$15, and sometimes lower in the \emph{J} band) restrict the use of narrow spectral features for classification. However, even at this resolution, there are still significant differences between the spectra of a low-gravity young stellar object and a field dwarf which can be studied. The triangular shape of the \emph{H} band, caused by deep H$_{2}$O absorption on either side of the sharp peak located between 1.68 and 1.70~$\mu$m in young objects, as opposed to a plateau in the spectra of field dwarfs, has been used as a signature of youth and membership in several studies of young brown dwarfs \citep[see, for example,][]{Allers2007,Lucas2001}. There are strong water absorption bands on both sides of the peak also in the \emph{K} band spectrum of young brown dwarfs. In the \emph{J}-band, H$_{2}$O absorption is also present at both extremes of the band for young brown dwarfs, but not for field dwarfs. Another good gravity indicator in the near-IR spectrum is the Na I absorption (present at 1.14 and 2.2~$\mu$m), which is very deep for field dwarfs, but not in young objects. Our spectral classification method relies on the comparison of the candidate spectra with those of young optically classified objects members of other star-forming regions of similar ages, which are used as standards. A numerical spectral fitting procedure was developed, which makes the simultaneous determination of spectral type and reddening possible. We also took spectra of sources outside the substellar selection limit we imposed in the colour-colour diagram (see Sect.~\ref{select:cmd}), to ensure that our selection criteria are not too stringent, and indeed all of those objects turned out to be stars with no water absorption features.
 
   \begin{table}
    \begin{minipage}[t]{\linewidth}
   \caption{Spectral type and A$_{\emph{V}}$ determined through numerical spectral fitting.}           
   \centering             
    \renewcommand{\footnoterule}{}
       \begin{tabular}{l c c c}       
   \hline           
   \hline
 CFHTWIR-Oph & \multicolumn{2}{c}{Spectral Type}  &  {A$_{\emph{V}}$ (mag)}  \\
  ~                  & Num. Fit. & H$_{2}$O Index           &     \\
   \hline                        
4     & M6.50$^{+0.25}_{-0.25}$ &  6.41$\pm$1 &    2.5$^{+0.1}_{-0.1}$     	\\
34  & M8.25$^{+0.5}_{-1.0}$ &  7.98$\pm$1 &    9.70$^{+0.7}_{-0.4}$    \\                 
47     & M7.50$^{+0.5}_{-0.5}$ &  7.09$\pm$1 &    5.6$^{+0.3}_{-0.2}$    \\
57     & M7.25$^{+0.75}_{-2.0}$ &  7.12$\pm$1 &    6.10$^{+1.7}_{-0.6}$    \\
62      & M5.50$^{+0.5}_{-1.25}$ &  4.01$\pm$1 &   9.90$^{+1.8}_{-0.9}$  \\
96  & M8.25$^{+0.25}_{-1.0}$ &  7.51$\pm$1 &    1.10$^{+0.7}_{-0.1}$   \\
106      & M6.50$^{+1.25}_{-1.0}$ &  6.42$\pm$1 &    4.9$^{+0.6}_{-0.7}$   \\
 \hline                                 
   \label{table:4} 
   \end{tabular}
   \end{minipage}
   \end{table}

\subsection{Numerical spectral fitting}
\label{specfit}
The procedure consists in comparing each candidate spectrum to a grid of near-IR, low-resolution template spectra of young stars and brown dwarfs with spectral types determined in the optical, reddened in even steps of A$_{\emph{V}}$. The comparison spectra are of members of the young ($\lesssim$2~Myr) star-forming regions IC~348 \citep{Luhman2003b} and Taurus \citep{Briceno2002,Luhman2004} dereddened by their extinction published values (typically A$_{\emph{V}}$~$\lesssim$~1), with spectral types ranging from M4 to M9.5. By combining the spectral types, half and quarter sub-classes were constructed. The reddening law of \citet{Fitzpatrick1999} was used to progressively redden the template spectra by steps of 0.1~A$_{\emph{V}}$. The fit was performed across the complete usable wavelength range of the spectrum  (1$-$1.34, 1.48$-$1.8, and 2$-$2.45~$\mu$m), i.e., excluding only the regions dominated by telluric absorption and the extremes of the spectral range where the quality of the data is poorer. It was assumed that all template spectra have approximately the same error. For the candidate spectra, the r.m.s. of the difference between the original spectrum and the smoothed spectrum (using a 15~pix boxcar) was taken as an estimation of the errors.  Figure~\ref{chi2fit} shows the contours for the variation of $\chi$$^{2}$ with A$_{\emph{V}}$ and spectral type for one of the candidates. The solid line contour represents the 1 sigma confidence interval, and the dashed lines indicate the A$_{\emph{V}}$ and spectral type limiting values at the point of the grid closest to the 1 sigma contour, which are taken as the standard deviation for each parameter from the minimum. The dotted contours are the 1.6, 2, 2.6, and 3 sigma levels, successively from the minimum. The right-hand panel shows the resulting best fit, CFHTWIR-Oph 34 is best-fitted by an M8.25 brown dwarf (an average spectra between an M8 and M9 young brown dwarfs) and an A$_{\emph{V}}$ of 9.7 magnitudes. This procedure was applied to all the candidate spectra. 

 \begin{figure*}
   \centering
 \includegraphics[width=\linewidth]{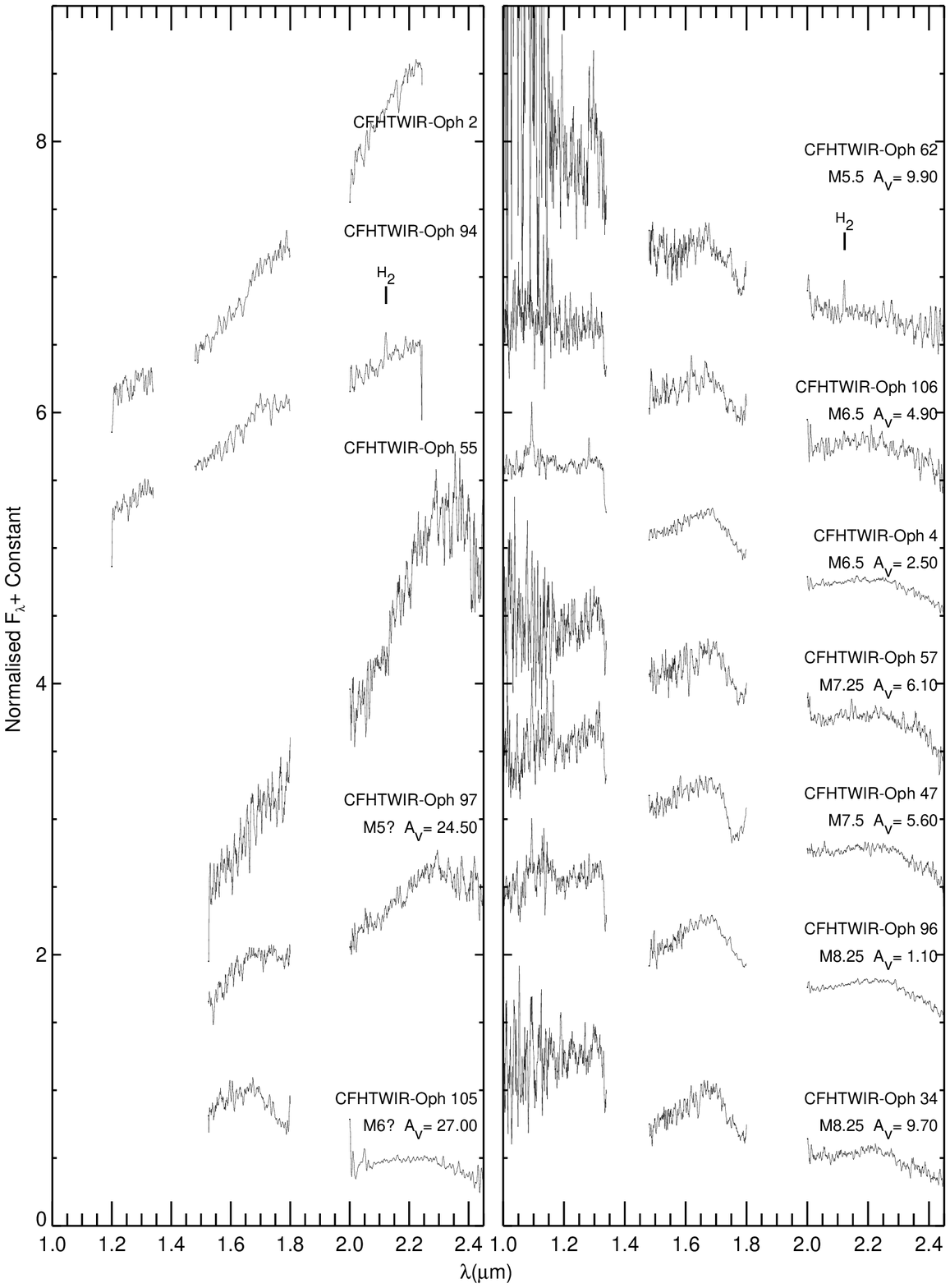}
   \caption{SofI/NTT and NICS/TNG  low-resolution spectra of the observed candidate (left panel) and confirmed (right panel) members in $\rho$~Oph. The spectra on the right panel are corrected for extinction with the values found through the numerical fitting procedure, which are shown together with the spectral type derived for the best-fit solution.}
   \label{allspec}
    \end{figure*}

\subsection{Spectral classification: comments on individual sources}
\label{comspec}
\subsubsection{Contaminant field stars}
Water vapour absorption could not be detected in the spectra of four candidates, and only a limit to the spectral type could be set, i.e., they have a spectral type earlier than $\sim$M-type. Given their faint IR magnitudes, they are therefore inconsistent with being young and members of the cluster and are excluded as background contaminants (listed in Table~\ref{cont}).

\subsubsection{CFHTWIR-Oph~4, 34, 47, 57, 62, 96, 106}  
The sources CFHTWIR-Oph~4, 34, 47, 57, 62, 96, and 106 have spectral types and extinction values derived from the numerical procedure, with spectral types ranging from M5.5 to M8.25, and A$_{\emph{V}}$ from a 1 to 10 magnitudes. The results are summarized in Table~\ref{table:4}, while the dereddened spectra are displayed in Fig.~\ref{allspec}. 

\subsubsection{GY~201}
The source GY~201 was observed with the blue and red grisms of SofI/NTT, and could neither be fitted with one of the templates in the grid, nor with comparison spectra from field dwarfs when the full spectral range was considered. If the fit was performed with only the part of the spectrum acquired with the red grism (from 1.5 to 2.5~$\mu$m), the fitting procedure converged for spectral type M5 and an A$_{\emph{V}}$ of 1.9 magnitudes. However, when the full spectrum was taken into account, no physical solution was found, with the resulting best fit indicating a negative value of A$_{\emph{V}}$. This disagreement can be explained if the object is an unresolved binary, where one of the components is an earlier type dwarf contributing to the blue part of the spectrum, and the other a late M-type dwarf dominating the red part of the spectrum and showing water vapour absorption features. Yet another plausible and more likely explanation is that the red part of its spectrum is contaminated by the emission from a nearby ($\le$10$\arcsec$) Class~I young member of $\rho$~Oph previously studied by many authors \citep[ISO~103, see for example,][]{Gutermuth2009,Padgett2008,Imanishi2001,Bontemps2001}. The source ISO~103 is very bright in the \emph{K} band, and is saturated in our WIRCam images. The source GY~201 was also previously associated with the cloud based on two near-IR studies \citep{Greene1992,Allen2002}, but was not detected or mentioned in any other study of the cloud, even if its location has been covered by the vast majority of the surveys. Its position on the various optical and near-IR CMDs presented in this paper indicates that this object could be a field dwarf, because its position is always bluewards of the models and other candidate members. It also lacks an IRAC or MIPS detection in the \emph{Spitzer} archive catalogues, which could be explained by the difficulty of separating its PSF from that of the neighbouring Class~I source, which is also very bright in the mid-IR. It has also not been detected by \citet{Gutermuth2009} or \citet{Padgett2008} in their two detailed \emph{Spitzer} studies of $\rho$~Oph. The nature of this object, and in particular its association to the cloud, remains therefore uncertain. Furthermore, in the review by \citet{Wilking2008}, the authors claim this object to be a Class~I source based on (unpublished) IRAC colours, a result which seems unlikely taking into account the results from the spectrum analysis in our work and its near-IR colours, and could be a mismatch between the IRAC detections and the existing literature catalogues, or IRAC photometry confusion caused by the neighbouring star. 

\subsubsection{CFHTWIR-Oph~2, 55, 94, 97, 105} 
The candidate members CFHTWIR-Oph~2 and 94 show a very red spectrum, without photospheric features required for a spectral classification, like clear water vapour absorption bands. The nature of these sources is further discussed in Sect.~\ref{properties}. Figure~\ref{allspec} shows the original spectra not dereddened, because they lack a classification. These objects were observed with a grism covering the entire near-IR spectrum (NICS/TNG), while CFHTWIR-Oph~55, 97, and 105 were observed with SofI/NTT using only the red grism, because they are too faint in the \emph{J} band. These spectra are also very red, though water vapour absorption can be seen in the spectra of CFHTWIR-Oph~97 and 105. These candidate members are classified as M5.5 and M6, respectively, though their classification is less reliable given that only a limited part of their spectra is available for the fit. The two other objects did not show clear absorption bands and have not been fitted. Their original undereddened spectra are also displayed in Fig.~\ref{allspec}, and their properties are discussed in detail in Sect.~\ref{properties}. 

\subsection{The H$_{2}$O Spectral Index}
For the seven candidate members with a spectral classification we compared the results from the numerical spectral fitting process with the H$_{2}$O spectral index defined by \citet{Allers2007}, which can be calculated as  $\langle F_{\lambda=1.550-1.560} /  F_{\lambda=1.492-1.502}\rangle$. \citet{Allers2007} derive a spectral type \emph{vs.} index relationship, which is independent of gravity and is valid for spectral types from M5 to L0, with an uncertainty of $\pm$~1 subtype. The index is computed for all our spectra after being dereddened by the A$_{\emph{V}}$ values in Table~\ref{table:4}, and the spectral types all agree with those determined by the fitting procedure within the uncertainties and are shown for comparison, further confirming the validity of our classification method.


\section{Properties of the spectroscopic sample}
\label{properties}

\subsection{Membership}
We used the compiled information for each candidate member observed spectroscopically to confirm their pre-main-sequence nature. For the seven sources that have determined spectral types and extinction values (Table~\ref{table:4}), we found an agreement with the spectra of young stellar objects with low gravity, and therefore identify in their spectra signatures of youth like the \emph{H}-band triangular shape. Additionally, all objects that have an optical counterpart (CFHTWIR-Oph~34, 47, 62, 96, and 106) show colours similar to those of the previously known brown dwarfs in $\rho$~Oph. In the mid-IR diagrams, CFHTWIR-Oph~2, 34, 55, 62, 94, 96, 97, and 105 show evidence of discs (Sect.~\ref{comp:spitzer}). 

Some of these objects have been previously associated with the cloud. The source CFHTWIR-Oph~62 was previously associated with the cloud from a comparison of near-IR photometric observations to models and the detection of infrared excess \citep{Rieke1990,Comeron1993,Greene1992}, and it was first observed spectroscopically by \citet{Wilking1999}, but lacked a high enough signal-to-noise to be studied. \citet{Cushing2000} observed the same object in the near-IR (NIRC/Keck~I), and claim its membership based on the detection of strong H$_{2}$ emission in the \emph{K}-band, though the authors mention it is not clear if the emission is associated with the object. H$_{2}$ emission is also detected in our SofI/NTT spectra, but the resolution of the spectrum is too low for an accurate velocity measurement (see also Sect.~\ref{outflow}). The spectral type derived by \citet{Cushing2000} is M4$\pm$1.3, which agrees within the errors, with the spectral type we found, M5.5$^{+0.5}_{-1.5}$. The sources CFHTWIR-Oph~34 and 96 have also been previously associated with the cloud based on the detection of near-IR excess \citep{Greene1992}, but lack a spectroscopic confirmation.  These three sources have mid-IR colours consistent with those expected for YSOs \citep{Wilking2008,Gutermuth2009}. Finally, CFHTWIR-Oph~34 and 62 have been classified as variable sources by \citet{Alvesdeoliveira2008}, which further supports their classification as members.

Despite the fact that the candidate members with very red spectra could not be classified, there is evidence they are members of the cluster. The sources CFHTWIR-Oph~55, 94, 97, and 105 have been previously associated with the cloud from IR and / or X-ray surveys \citep{Wilking2008}. The source CFHTWIR-Oph~2 is a new candidate member and shows colours consistent with those of a Class~II object (Fig.~\ref{spitzer}). 

\subsection{H$_{2}$ Outflow}
\label{outflow}
The source CFHTWIR-Oph~94 (other names are, for example, GY~312 or ISO~165) is a known member of $\rho$~Oph and has been extensively studied: \citet{Imanishi2001} detected both quiescent and flare X-ray emission, \citet{Natta2006} found it to be an actively accreting YSO, and \citet{Alvesdeoliveira2008} detected photometric variability consistent with changes in the surrounding disc or envelope. \citet{Bontemps2001} classified it as a Class~II object, i.e., with an IR excess and a spectral energy distribution (SED) which can be explained by models of YSOs surrounded by circumstellar discs. More recently, using \emph{Spitzer} data, \citet{Gutermuth2009} have classified it as a Class~I, given its strong IR excess. Our mid-IR colour-colour magnitude diagrams agree with the later classification. We found further evidence of the protostellar nature and therefore youth of this object. We detected a  H$_{2}$~1~-~0 S(0) emission (2.12~$\mu$m) in the spectrum of CFHTWIR-Oph~94 (Fig.~\ref{allspec}), a signature of a molecular outflow. Given the extreme red spectrum of this object, we cannot estimate its spectral type. Further observations are needed to determine the association of the outflow with the source \citep[see, for example, ][]{Bourke2005,Fernandez2005}, and also its mass.

\subsection{Candidate edge-on disc}
\label{candidate_disc}

   \begin{figure}
   \centering
 \includegraphics[width=\columnwidth]{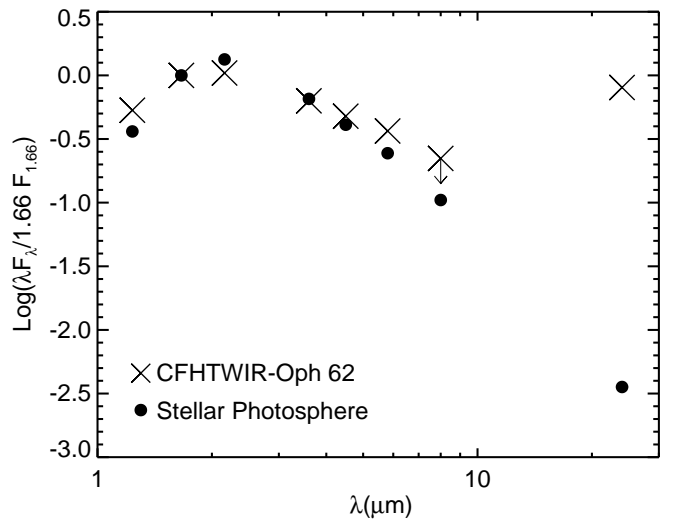}
   \caption{SED of a candidate edge-on disc in $\rho$~Oph. The SED of CFHTWIR-Oph~62 is compared to the SED of a known Class~III young stellar object in $\rho$~Oph with a similar spectral type, reddened by an extinction amount to match that of the candidate. Both sources are scaled at the \emph{H}-band flux. This object is underluminous at optical wavelengths, does not show an excess emission from the near-IR up to  8~$\mu$m, but has a large excess at 24~$\mu$m.}
   \label{transdisc}
    \end{figure}

We investigated the mid-IR colours of a candidate edge-on disc, CFHTWIR-Oph~62, which shows very red colours at 24~$\mu$m but is not present in the IRAC diagrams (the object is detected at 3.6, 4.5, and 5.8~$\mu$m, but has only an upper limit detection at 8~$\mu$m). We compared the colours of CFHTWIR-Oph~62 to those of WSB~50, a young stellar object member of $\rho$~Oph with a spectral type close to that of CFHTWIR-Oph~62, between M4.5 \citep{Wilking2005} and M4 \citep{Luhman1999}, and classified as a Class~III source, which should therefore show a \emph{photospheric} SED. The magnitudes are reddened to those of CFHTWIR-Oph~62 (with an A$_{\emph{V}}$ of 8., WSB~50 has an A$_{\emph{V}}$=1.9 magnitudes) with the reddening laws from \citet{Rieke1985} and \citet{Flaherty2007}. The near-IR magnitudes were taken from the 2MASS catalogue because WSB~50 is saturated in the CFHT/WIRCam images, and it is preferred to use a common photometric system. Figure~\ref{transdisc} shows the two SEDs, which nicely shows the characteristic SED of CFHTWIR-Oph~62. \citet{Cushing2000} found this object to be underluminous at optical wavelengths, which combined with the lack of IR excess up to 8~$\mu$m could be explained by the geometry of a nearly edge-on disc: at short wavelengths the disc is optically thick and acts as a natural coronograph (explaining the underluminosity when compared to members of similar T$_{eff}$, and why \citet{Cushing2000} concluded this object to be older), while at longer wavelengths the thermal emission of the disc dominates, causing the sharp rise in the SED \citep[see, for example][]{Sauter2009,Duchene2009}. That the flux at 24~$\mu$m is at the level of the \emph{photospheric} part of the SED further supports this scenario. Further observations and modelling are needed to understand and better characterise this complex young object. In particular, measurements at longer wavelengths than 24~$\mu$m can provide an important constraint in the nature of this source.  

\subsection{Temperatures and luminosities}

 \begin{figure*}
   \centering
 \includegraphics[width=\linewidth]{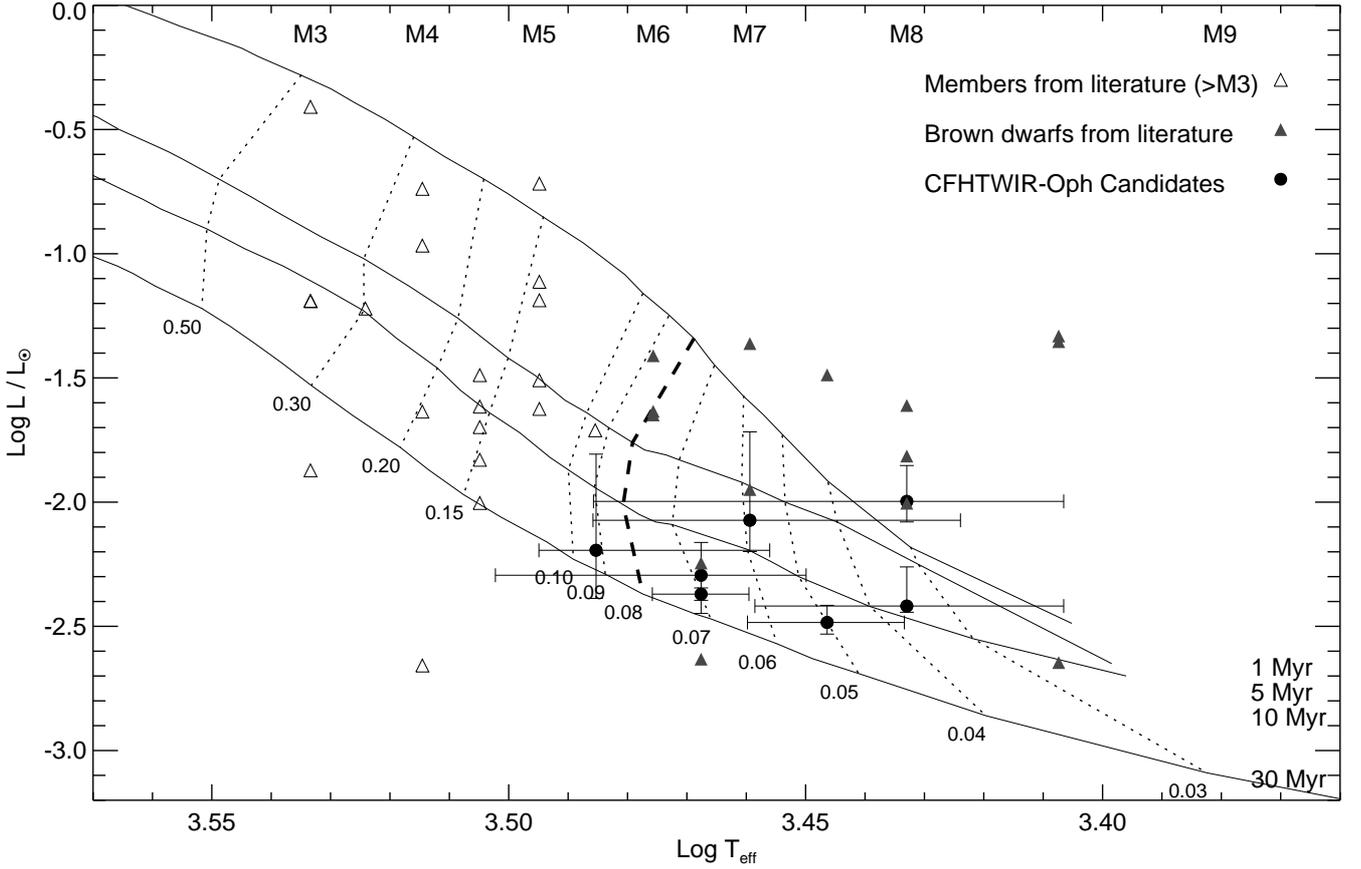}
   \caption{H-R diagram for $\rho$~Oph displaying the new brown dwarf candidates found in this study, together with previously known stellar and substellar members of the cloud. The models are the NextGen isochrones \protect{\citep{Baraffe1998}} shifted to the distance of the cluster (130~pc), for 1, 5, 10, and 30~Myr, and labelled with mass in units of M$_{\sun}$. }
   \label{hr}
    \end{figure*}

To place the new candidate members that were spectroscopically confirmed in the Hertzsprung-Russell (HR) diagram, which is commonly used to estimate ages and masses, we needed to derive their effective temperature and bolometric luminosity.  To convert spectral types to temperatures, the temperature scale from \citet{Luhman2003} was adopted, which is derived for young members of the star-forming region IC~348 ($\sim$2~Myr), and has provided consistent results when applied to other young star-forming regions \citep[for example,][]{Luhman2009}. The adopted errors for the temperature are the one sigma limits in spectral type from the numerical fitting procedure. For each candidate, the bolometric luminosity was calculated from the dereddened \emph{J} magnitude (using A$_{\emph{V}}$ derived from the numerical fit and the reddening law from \citet{Rieke1985}), applying the bolometric corrections for the respective spectral type (from \citet{Kenyon1995} for $<$M6 and \citet{Dahn2002} for $\ge$M6), and using a distance to the cloud of 130~pc. The errors were propagated to include the photometric error in \emph{J}, the one sigma errors in A$_{\emph{V}}$ from the spectral fitting, and by assuming an error of $\pm$10~pc in the distance to the cloud.  For comparison purposes, we compiled the previously known members of $\rho$~Oph with assigned spectral types later then $\gtrsim$~M3 from the literature. From the $\sim$300 objects associated with the cloud,  members that have a counterpart and are not saturated in the WIRCam/CFHT \emph{J}-band images were kept. Some of these objects are not part of the final WIRCam catalogue (see Sect.~\ref{data}), because they are saturated either in the \emph{H} or the \emph{K$_{s}$} bands. Only objects that have spectral types determined from spectroscopic surveys and extinction values published are included. The final compilation contains 36 young low-mass stars and brown dwarfs \citep{Luhman1997,Luhman1999,Wilking1999,Cushing2000,Natta2002,Wilking2005}. Bolometric luminosities and temperatures were derived in the same way as for the CFHTWIR-Oph candidates. The previously confirmed members and candidate members from this study were placed in the HR diagram (Fig.~\ref{hr}) and compared to theoretical evolutionary models \citep[NextGen, because our candidates have T$_{\emph{eff}}$ $>$ 2500~\emph{K},][]{Baraffe1998}. The 1, 5, 10, and 30~Myr isochrones are shown, labelled with mass in units of M$_{\sun}$. We adopted the 0.08~M$_{\sun}$ mass track as the stellar / substellar boundary that corresponds to spectral types $\sim$M6.25 to M6.5 for a young member of $\rho$~Oph with an age of 1 to 2~Myr. In our sample, we find that CFHTWIR-Oph~62 is a very low-mass star, and the other sources are six new brown dwarfs of $\rho$~Oph (CFHTWIR-Oph~4, 34, 47, 57, 96, 106). 

\begin{figure}
   \centering
 \includegraphics[width=\columnwidth]{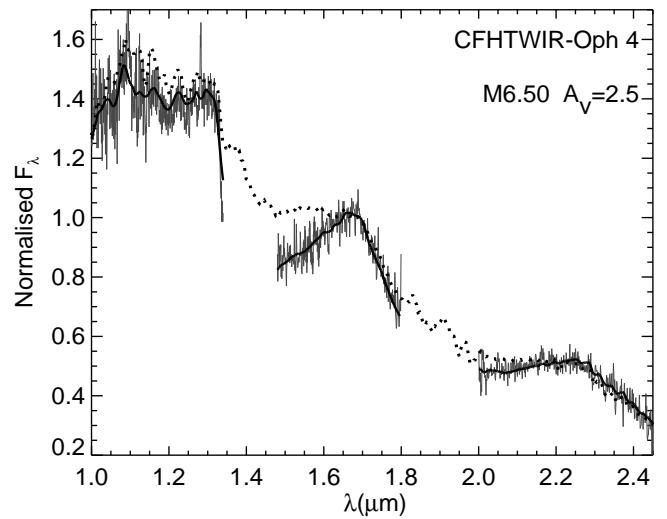}
   \caption{Dereddened spectrum of CFHTWIR-Oph~4 together with the best fit obtained. The spectrum shows a very good match to that of an intermediate spectrum between an M6 and M7 template young brown dwarfs (solid line) and clear differences to the spectrum of the field dwarf 2MASS J13272391+0946446 \protect{\citep{Burgasser2004}} with a comparable spectral type (dotted line).}
   \label{youngfield}
    \end{figure}

A large spread in ages is seen in the HR diagram, which goes from $<$1 to $\sim$10~Myr, and with some objects laying already closer to the 30~Myr isochrone. The estimated age for $\rho$~Oph is of 0.3~Myr in the core \citep{Greene1995,Luhman1999}, and 1-5~Myr in the surrounding regions \citep{Bouvier1992,Martin1998,Wilking2005}. The star formation history of this cluster is thought to be connected to that of the Sco-Cen OB association, with two different episodes of star formation taking place, one caused by a supernova 1 to 1.5~Myrs ago, and the other happening in parallel to the formation of Upper Scorpius $\sim$5~Myrs ago, caused by an expanding shell from the Upper Centaurus-Lupus OB subgroup  \citep[see][and references therein for a review]{Wilking2008}. These ideas are still debated, but would mean that a range of ages is therefore expected in the HR diagram. Taking into account the typical uncertainties involved in the temperature and A$_{\emph{V}}$ determination, that would explain the position of the members of the cluster from above the 1~Myr isochrone up to the 10~Myr models. Most of the members in the HR diagram are consistent with this age estimate, which is also the case for our candidates CFHTWIR-Oph~34, 57, and 96. The other CFHTWIR-Oph candidates have luminosities that suggest an age older than 10~Myr up to 30~Myr. We already mentioned, however, that CFHTWIR-Oph~62 is underluminous \citep{Cushing2000}, most probably due to observations being done through scattered light from a surrounding disc (Sect.~\ref{candidate_disc}). The old ages implied in the diagram for CFHTWIR-Oph~4, 47, and 106, do not seem plausible if we assume them to be members of the cluster. The source CFHTWIR-Oph~106 shows mid-IR colours consistent with those of young objects surrounded by discs, and it is possible that it is observed through scattered light, which would explain its lower luminosity. Both CFHTWIR-Oph~4 and 47 do not show a signature of IR excess and could be more evolved young objects. As mentioned in Sect.~\ref{specfit}, their spectra are well fitted by those of young objects, showing distinctive features of youth, in particular the triangular shape of the \emph{H}-band. Figure~\ref{youngfield} shows the dereddened spectrum of CFHTWIR-Oph~4 together with the best-fit spectrum from the young grid of templates, and a comparison spectrum of a field dwarf of the same spectral type, putting into evidence the pronounced and broad water absorption bands associated with low surface gravity objects \citep{Kirkpatrick2006}. The same check was done for all our classified spectra. Furthermore, three other brown dwarfs taken from the literature (CRBR~31, GY~11, GY~141) fall into the same part of the HR diagram. We do not find any relation between the different positions of brown dwarfs in the HR diagram (younger or older than the 10~Myr isochrone) and their positions on sky in relation to the cluster's core, for example. Nor do we find a relation between their position in the HR diagram and their SED class assigned from mid-IR colours. In particular, all but one of the previously known brown dwarfs in the cloud with an assigned SED class are Class~II objects.

   \begin{table*}
   \begin{minipage}[t]{\linewidth}
   \caption{Candidate brown dwarfs in $\rho$~Ophiuchus known to date.}           
   \centering             
    \renewcommand{\footnoterule}{}
       \begin{tabular}{l r c c c c}       
   \hline           
   \hline
Name  &  \emph{J}\footnote{PSF photometry magnitude from WIRCam/CFHT survey, unless noted otherwise.} & A$_{\emph{V}}$ &Sp. Type & T$_{\emph{eff}}$\footnote{From the SpT- T$_{\emph{eff}}$ relation of \citet{Luhman2003}.} & Ref.\footnote{Spectral types determined from the following studies: 1.~\citet{Luhman1997}; 2.~\citet{Wilking1999}; 3.~\citet{Luhman1999}; 4.~\citet{Cushing2000}; 5.~\citet{Natta2002}; 6.~\citet{Wilking2005}.}  \\
~ & ~(mag) &  (mag) & ~ & (\emph{K}) &   ~ \\
   \hline                    	    
CFHTWIR-Oph~4    &  14.88  & 2.5  &   M6.50   &	2935$^{+55}_{-55}$  & this work  \\ 
CFHTWIR-Oph~34   &  15.97  & 9.70 &   M8.25   &	2710$^{+550}_{-170}$  & this work  \\
CFHTWIR-Oph~47   &  15.94  & 5.6  &   M7.50   &	2795$^{+85}_{-85}$  & this work  \\
CFHTWIR-Oph~57   &  15.06  & 6.10 &   M7.25   &	2880$^{+170}_{-245}$  & this work  \\
CFHTWIR-Oph~96~  &  14.60  & 1.10 &   M8.25   &  2710$^{+310}_{-170}$  & this work  \\
CFHTWIR-Oph~106~  &  15.36  & 4.9  &   M6.50   &	2935$^{+225}_{-122}$  & this work  \\
CRBR~14	  & 14.79 & 10.0  & M7.5(M5.5,M7)  &   2795	  &  2,3,5  \\ 
CRBR~31   & 16.28 & 8.6   &  M6.7	    & 	2935	  &  4 \\ 
GY~3   	  & 12.27 & 0.0   &  M8(M7.5)	    & 	2710	  &  6,5 \\
GY~5   	  & 12.44 & 2.8   &  M7(M6)	    & 	2880	  &  2,5 \\
GY~10  	  & 15.52 & 14.0  &  M8.5(M6.5)      & 	2555	  &  2,3 \\
GY~11  	  & 16.17 & 4.8   &  M6.5(M8.5)      & 	2935	  &  2,5 \\
GY~64  	  & 16.35 & 11.0  &  M8 	    & 	2710	  &  2 \\
GY~141    & 15.05 & 0.7   &  M8.5(M8) 	    & 	2555	  &  1,4 \\ 
GY~202    & 16.79 & 13.0  &  M7(M6.5)	    & 	2880	  &  2,3 \\
GY~204    & 12.47 & 0.5   &  M6 	    & 	2990	  &  5 \\
GY~264    & 12.78 & 0.0   &  M8    & 	2710	  &  6 \\
GY~310    & 13.23 & 5.7   &  M8.5(M7,M6)     &   2555	  &  2,3,5  \\
GY~350    & 13.74 & 7.0   &  M6 	    & 	2990	  &  5 \\ 
oph-160   & 14.05 & 6.0   &  M6  	    & 	2990      &  5 \\ 
oph-193   & 13.61\footnote{This object is not part of the WIRCam surveyed region. Photometry is from 2MASS public catalogue.} & 7.5 	&  M6   	&	2990 		&  5 \\
  \hline                            
   \label{bd} 
   \end{tabular}
   \end{minipage}
   \end{table*}
 
The old ages implied by the isochrones could instead be related to the several sources of error associated with this diagram, like the less reliable photometry for sources located in high nebulosity regions (abundant in $\rho$~Oph), near-IR variability in YSOs (\citet{Alvesdeoliveira2008} detected near-IR variations as large as 0.5 magnitudes for objects plotted in the HR diagram), unresolved binaries, the large difference in estimated distances to the cloud, or the possibility that some objects may be seen through scattered light. Although the individual contribution of these uncertainties can be quantified, it is not possible to have a clear picture of the net effect when one or more of the mentioned problems are involved. Furthermore, recent results in modelling of young brown dwarfs \citep{Baraffe2009} suggest that episodic strong accretion might explain the observed spread in HR diagrams at ages of a few Myr years, a scenario supported by recent observations of protostars, some of which were carried out in $\rho$~Oph \citep{Enoch2009}. According to these results, even after accretion has halted, young low-mass objets can keep a memory from these strong accretion events, altering the expected path in their contraction along the Hayashi track, and therefore their position in the HR diagram.

Another possible explanation could also be that some of these brown dwarfs lie behind $\rho$~Oph and are instead members of Upper Sco, which would mean the luminosity is underestimated in the HR diagram presented. If we compute the bolometric luminosities using a distance of 165~pc instead (approximate boundary of Upper Sco), all sources are within the 10~Myr isochrone or younger in the HR diagram. Though it is unlikely this is the case for all the sources, it is possible that some of the brown dwarfs associated with $\rho$~Oph could rather be Upper Sco members.

Given these uncertainties, we have therefore not assigned a definite age or mass to the newly confirmed brown dwarfs and the very low-mass star discovered in this work. We can claim though, based on their position in the HR diagram relative to the other members of the cloud, that they have ages and luminosities which agree with those of the known substellar members. From their location in the HR diagram, these new candidates indeed appear to be amongst the lowest mass objects of the cluster. 

\subsubsection{$\rho$~Ophiuchi: census update of the substellar population}

We compiled from previous studies a list of spectroscopically confirmed members of $\rho$~Oph with spectral types later than $\sim$M6 that are therefore likely to be brown dwarfs (according to the evolutionary models of \citet{Baraffe1998}) and present them in Table~\ref{bd} together with the six new brown dwarfs found in this work. All surveys were conducted in the main cloud, L1688. All but one (not in the WIRCam/CFHT survey coverage) of the brown dwarfs are plotted in the HR diagram in Fig.~\ref{hr}.  We did not include in this list the brown dwarfs discovered by \citet{Allers2007} in a region to the north west of the central cloud, because it has been suggested that two of the three are  associated with an older population of Sco-Cen \citep{Luhman2007a} and therefore their membership to the $\rho$~Ophiuchi cloud complex has not been confirmed \citep[see also, ][]{Close2007}. We did not include in this list the T2 dwarf found by \citet{Marsh2009} either, because the main argument used to claim membership to $\rho$~Oph relies on the distance determination using the apparent \emph{K}-band magnitude of the object, which we claim might be wrong. This list provides an updated census of the substellar members of $\rho$~Oph known to date.


\section{Conclusion}
\label{conclusion}

We identify 110 substellar candidate members of $\rho$~Ophiuchi from a deep, near-IR photometric survey, from which 80 were not previously associated with the cloud. By extensive use of archive multi-wavelength data, we find evidence of mid-IR excess for 27\% of the candidates and a variability behaviour consistent with that of YSOs for 15\%, further supporting the membership of these candidates. 

We started a spectroscopic follow-up of the substellar candidate members, and present the first results for 16 sources. We identify six new members of $\rho$~Ophiuchi with spectral types ranging from $\sim$M6.5 to $\sim$M8.25, and classify them as new confirmed brown dwarfs according to the evolutionary models of \citet{Baraffe1998}. We confirm the spectral type derived by \citet{Cushing2000} for a previously known very low-mass star close to the substellar limit, and based on the SED constructed from optical to mid-IR photometry, we report the discovery of a candidate edge-on disc around this star. We cannot derive accurate spectral types for five sources which have extremely red spectra. Two of these show water absorption features and are classified with spectral types M5 and M6. However, since they lack a \emph{J}-band spectra and given the poor fit they remain as candidate members. The remaining three sources could be T~Tauri star members of the cluster, because they show strong mid-IR excess and one of them is emitting in X-rays. We found signatures of outflow activity in two of the sources studied spectroscopically where H$_{2}$~1~-~0 S(0) emission (2.12~$\mu$m) was detected. Four sources out of the 16 were found to be contaminant field dwarfs.

\begin{acknowledgements}
We thank Dr. Kevin Luhman for his useful comments as a referee, and for providing template spectra of young stars and brown dwarfs. We thank Dr. Ignazio Pillitteri, Dr. Ettore Flaccomio, and collaborators from the DROXO team for providing some of their results prior to publication. We thank the QSO team at CFHT for their efficient work at the telescope and the data pre-reduction as well as the Terapix group at IAP for the image reduction. This work is based in part on data products produced and image reduction processes conducted at TERAPIX. This research has made use of the NASA/ IPAC Infrared Science Archive, which is operated by the Jet Propulsion Laboratory, California Institute of Technology, under contract with the National Aeronautics and Space Administration. This research has also made use of the SIMBAD database, operated at CDS, Strasbourg, France.
 
\end{acknowledgements}


\bibliographystyle{aa}
\bibliography{alvesdeoliveira}

\begin{thebibliography}{111}
\expandafter\ifx\csname natexlab\endcsname\relax\def\natexlab#1{#1}\fi

\bibitem[{{Allen} {et~al.}(2004){Allen}, {Calvet}, {D'Alessio}, {Merin},
  {Hartmann}, {Megeath}, {Gutermuth}, {Muzerolle}, {Pipher}, {Myers}, \&
  {Fazio}}]{Allen2004}
{Allen}, L.~E., {Calvet}, N., {D'Alessio}, P., {et~al.} 2004, \apjs, 154, 363

\bibitem[{{Allen} {et~al.}(2002){Allen}, {Myers}, {Di Francesco}, {Mathieu},
  {Chen}, \& {Young}}]{Allen2002}
{Allen}, L.~E., {Myers}, P.~C., {Di Francesco}, J., {et~al.} 2002, \apj, 566,
  993

\bibitem[{{Allers} {et~al.}(2007){Allers}, {Jaffe}, {Luhman}, {Liu}, {Wilson},
  {Skrutskie}, {Nelson}, {Peterson}, {Smith}, \& {Cushing}}]{Allers2007}
{Allers}, K.~N., {Jaffe}, D.~T., {Luhman}, K.~L., {et~al.} 2007, \apj, 657, 511

\bibitem[{{Alves de Oliveira} \& {Casali}(2008)}]{Alvesdeoliveira2008}
{Alves de Oliveira}, C. \& {Casali}, M. 2008, \aap, 485, 155

\bibitem[{{Andersen} {et~al.}(2008){Andersen}, {Meyer}, {Greissl}, \&
  {Aversa}}]{Andersen2008}
{Andersen}, M., {Meyer}, M.~R., {Greissl}, J., \& {Aversa}, A. 2008, \apjl,
  683, L183

\bibitem[{{Baba} {et~al.}(2002){Baba}, {Yasuda}, {Ichikawa}, {Yagi}, {Iwamoto},
  {Takata}, {Horaguchi}, {Taga}, {Watanabe}, {Ozawa}, \& {Hamabe}}]{Baba2002}
{Baba}, H., {Yasuda}, N., {Ichikawa}, S., {et~al.} 2002, in Astronomical
  Society of the Pacific Conference Series, Vol. 281, Astronomical Data
  Analysis Software and Systems XI, ed. {D.~A.~Bohlender, D.~Durand, \&
  T.~H.~Handley}, 298--+

\bibitem[{{Baraffe} {et~al.}(1998){Baraffe}, {Chabrier}, {Allard}, \&
  {Hauschildt}}]{Baraffe1998}
{Baraffe}, I., {Chabrier}, G., {Allard}, F., \& {Hauschildt}, P.~H. 1998, \aap,
  337, 403

\bibitem[{{Baraffe} {et~al.}(2009){Baraffe}, {Chabrier}, \&
  {Gallardo}}]{Baraffe2009}
{Baraffe}, I., {Chabrier}, G., \& {Gallardo}, J. 2009, \apjl, 702, L27

\bibitem[{{Barsony} {et~al.}(1997){Barsony}, {Kenyon}, {Lada}, \&
  {Teuben}}]{Barsony1997}
{Barsony}, M., {Kenyon}, S.~J., {Lada}, E.~A., \& {Teuben}, P.~J. 1997, \apjs,
  112, 109

\bibitem[{{Bertin} \& {Arnouts}(1996)}]{Bertin1996}
{Bertin}, E. \& {Arnouts}, S. 1996, \aaps, 117, 393

\bibitem[{{Bihain} {et~al.}(2009){Bihain}, {Rebolo}, {Zapatero Osorio},
  {B{\'e}jar}, {Vill{\'o}-P{\'e}rez}, {D{\'{\i}}az-S{\'a}nchez},
  {P{\'e}rez-Garrido}, {Caballero}, {Bailer-Jones}, {Barrado y Navascu{\'e}s},
  {Eisl{\"o}ffel}, {Forveille}, {Goldman}, {Henning}, {Mart{\'{\i}}n}, \&
  {Mundt}}]{Bihain2009}
{Bihain}, G., {Rebolo}, R., {Zapatero Osorio}, M.~R., {et~al.} 2009, \aap, 506,
  1169

\bibitem[{{Bonnell} {et~al.}(2007){Bonnell}, {Larson}, \&
  {Zinnecker}}]{Bonnell2007}
{Bonnell}, I.~A., {Larson}, R.~B., \& {Zinnecker}, H. 2007, Protostars and
  Planets V, 149

\bibitem[{{Bontemps} {et~al.}(2001){Bontemps}, {Andr{\'e}}, {Kaas}, {Nordh},
  {Olofsson}, {Huldtgren}, {Abergel}, {Blommaert}, {Boulanger}, {Burgdorf},
  {Cesarsky}, {Cesarsky}, {Copet}, {Davies}, {Falgarone}, {Lagache},
  {Montmerle}, {P{\'e}rault}, {Persi}, {Prusti}, {Puget}, \&
  {Sibille}}]{Bontemps2001}
{Bontemps}, S., {Andr{\'e}}, P., {Kaas}, A.~A., {et~al.} 2001, \aap, 372, 173

\bibitem[{{Bourke} {et~al.}(2005){Bourke}, {Crapsi}, {Myers}, {Evans},
  {Wilner}, {Huard}, {J{\o}rgensen}, \& {Young}}]{Bourke2005}
{Bourke}, T.~L., {Crapsi}, A., {Myers}, P.~C., {et~al.} 2005, \apjl, 633, L129

\bibitem[{{Bouvier} \& {Appenzeller}(1992)}]{Bouvier1992}
{Bouvier}, J. \& {Appenzeller}, I. 1992, \aaps, 92, 481

\bibitem[{{Bouy} {et~al.}(2009{\natexlab{a}}){Bouy}, {Hu{\'e}lamo}, {Barrado Y
  Navascu{\'e}s}, {Mart{\'{\i}}n}, {Petr-Gotzens}, {Kolb}, {Marchetti},
  {Morales-Calder{\'o}n}, {Bayo}, {Artigau}, {Hartung}, {Marchis}, {Tamura},
  {Sterzik}, {K{\"o}hler}, {Ivanov}, \& {N{\"u}rnberger}}]{Bouy2009b}
{Bouy}, H., {Hu{\'e}lamo}, N., {Barrado Y Navascu{\'e}s}, D., {et~al.}
  2009{\natexlab{a}}, \aap, 504, 199

\bibitem[{{Bouy} {et~al.}(2009{\natexlab{b}}){Bouy}, {Hu{\'e}lamo},
  {Mart{\'{\i}}n}, {Marchis}, {Barrado Y Navascu{\'e}s}, {Kolb}, {Marchetti},
  {Petr-Gotzens}, {Sterzik}, {Ivanov}, {K{\"o}hler}, \&
  {N{\"u}rnberger}}]{Bouy2009a}
{Bouy}, H., {Hu{\'e}lamo}, N., {Mart{\'{\i}}n}, E.~L., {et~al.}
  2009{\natexlab{b}}, \aap, 493, 931

\bibitem[{{Brice{\~n}o} {et~al.}(2002){Brice{\~n}o}, {Luhman}, {Hartmann},
  {Stauffer}, \& {Kirkpatrick}}]{Briceno2002}
{Brice{\~n}o}, C., {Luhman}, K.~L., {Hartmann}, L., {Stauffer}, J.~R., \&
  {Kirkpatrick}, J.~D. 2002, \apj, 580, 317

\bibitem[{{Burgasser} {et~al.}(2007){Burgasser}, {Looper}, {Kirkpatrick}, \&
  {Liu}}]{Burgasser2007}
{Burgasser}, A.~J., {Looper}, D.~L., {Kirkpatrick}, J.~D., \& {Liu}, M.~C.
  2007, \apj, 658, 557

\bibitem[{{Burgasser} {et~al.}(2004){Burgasser}, {McElwain}, {Kirkpatrick},
  {Cruz}, {Tinney}, \& {Reid}}]{Burgasser2004}
{Burgasser}, A.~J., {McElwain}, M.~W., {Kirkpatrick}, J.~D., {et~al.} 2004,
  \aj, 127, 2856

\bibitem[{{Burgess} {et~al.}(2009){Burgess}, {Moraux}, {Bouvier}, {Marmo},
  {Albert}, \& {Bouy}}]{Burgess2009}
{Burgess}, A.~S.~M., {Moraux}, E., {Bouvier}, J., {et~al.} 2009, ArXiv e-prints

\bibitem[{{Chabrier} {et~al.}(2000){Chabrier}, {Baraffe}, {Allard}, \&
  {Hauschildt}}]{Chabrier2000}
{Chabrier}, G., {Baraffe}, I., {Allard}, F., \& {Hauschildt}, P. 2000, \apj,
  542, 464

\bibitem[{{Close} {et~al.}(2007){Close}, {Zuckerman}, {Song}, {Barman},
  {Marois}, {Rice}, {Siegler}, {Macintosh}, {Becklin}, {Campbell}, {Lyke},
  {Conrad}, \& {Le Mignant}}]{Close2007}
{Close}, L.~M., {Zuckerman}, B., {Song}, I., {et~al.} 2007, \apj, 660, 1492

\bibitem[{{Comeron} {et~al.}(1993){Comeron}, {Rieke}, {Burrows}, \&
  {Rieke}}]{Comeron1993}
{Comeron}, F., {Rieke}, G.~H., {Burrows}, A., \& {Rieke}, M.~J. 1993, \apj,
  416, 185

\bibitem[{{Cuillandre} {et~al.}(2004){Cuillandre}, {Magnier}, {Isani}, {Sabin},
  {Knight}, {Kras}, \& {Lai}}]{Cuillandre2004}
{Cuillandre}, J., {Magnier}, E.~A., {Isani}, S., {et~al.} 2004, in Astrophysics
  and Space Science Library, Vol. 300, Scientific Detectors for Astronomy, The
  Beginning of a New Era, ed. {P.~Amico, J.~W.~Beletic, \& J.~E.~Belectic},
  287--298

\bibitem[{{Cushing} {et~al.}(2000){Cushing}, {Tokunaga}, \&
  {Kobayashi}}]{Cushing2000}
{Cushing}, M.~C., {Tokunaga}, A.~T., \& {Kobayashi}, N. 2000, \aj, 119, 3019

\bibitem[{{Dahn} {et~al.}(2002){Dahn}, {Harris}, {Vrba}, {Guetter}, {Canzian},
  {Henden}, {Levine}, {Luginbuhl}, {Monet}, {Monet}, {Pier}, {Stone}, {Walker},
  {Burgasser}, {Gizis}, {Kirkpatrick}, {Liebert}, \& {Reid}}]{Dahn2002}
{Dahn}, C.~C., {Harris}, H.~C., {Vrba}, F.~J., {et~al.} 2002, \aj, 124, 1170

\bibitem[{{de Bruijne} {et~al.}(1997){de Bruijne}, {Hoogerwerf}, {Brown},
  {Aguilar}, \& {de Zeeuw}}]{deBruijne1997}
{de Bruijne}, J.~H.~J., {Hoogerwerf}, R., {Brown}, A.~G.~A., {Aguilar}, L.~A.,
  \& {de Zeeuw}, P.~T. 1997, in ESA Special Publication, Vol. 402, Hipparcos -
  Venice '97, 575--578

\bibitem[{{de Geus} {et~al.}(1989){de Geus}, {de Zeeuw}, \& {Lub}}]{deGeus1989}
{de Geus}, E.~J., {de Zeeuw}, P.~T., \& {Lub}, J. 1989, \aap, 216, 44

\bibitem[{{Duchene} {et~al.}(2009){Duchene}, {McCabe}, {Pinte}, {Stapelfeldt},
  {Menard}, {Duvert}, {Ghez}, {Maness}, {Bouy}, {Barrado y Navascues},
  {Morales-Calderon}, {Wolf}, {Padgett}, {Brooke}, \&
  {Noriega-Crespo}}]{Duchene2009}
{Duchene}, G., {McCabe}, C., {Pinte}, C., {et~al.} 2009, ArXiv e-prints

\bibitem[{{Enoch} {et~al.}(2009){Enoch}, {Evans}, {Sargent}, \&
  {Glenn}}]{Enoch2009}
{Enoch}, M.~L., {Evans}, N.~J., {Sargent}, A.~I., \& {Glenn}, J. 2009, \apj,
  692, 973

\bibitem[{{Evans} \& {c2d Team}(2005)}]{Evans2005}
{Evans}, N.~J. \& {c2d Team}. 2005, in Bulletin of the American Astronomical
  Society, Vol.~37, Bulletin of the American Astronomical Society, 1323--+

\bibitem[{{Evans} {et~al.}(2003){Evans}, {Allen}, {Blake}, {Boogert}, {Bourke},
  {Harvey}, {Kessler}, {Koerner}, {Lee}, {Mundy}, {Myers}, {Padgett},
  {Pontoppidan}, {Sargent}, {Stapelfeldt}, {van Dishoeck}, {Young}, \&
  {Young}}]{Evans2003}
{Evans}, II, N.~J., {Allen}, L.~E., {Blake}, G.~A., {et~al.} 2003, \pasp, 115,
  965

\bibitem[{{Fazio} {et~al.}(2004){Fazio}, {Hora}, {Allen}, {Ashby}, {Barmby},
  {Deutsch}, {Huang}, {Kleiner}, {Marengo}, {Megeath}, {Melnick}, {Pahre},
  {Patten}, {Polizotti}, {Smith}, {Taylor}, {Wang}, {Willner}, {Hoffmann},
  {Pipher}, {Forrest}, {McMurty}, {McCreight}, {McKelvey}, {McMurray}, {Koch},
  {Moseley}, {Arendt}, {Mentzell}, {Marx}, {Losch}, {Mayman}, {Eichhorn},
  {Krebs}, {Jhabvala}, {Gezari}, {Fixsen}, {Flores}, {Shakoorzadeh}, {Jungo},
  {Hakun}, {Workman}, {Karpati}, {Kichak}, {Whitley}, {Mann}, {Tollestrup},
  {Eisenhardt}, {Stern}, {Gorjian}, {Bhattacharya}, {Carey}, {Nelson},
  {Glaccum}, {Lacy}, {Lowrance}, {Laine}, {Reach}, {Stauffer}, {Surace},
  {Wilson}, {Wright}, {Hoffman}, {Domingo}, \& {Cohen}}]{Fazio2004}
{Fazio}, G.~G., {Hora}, J.~L., {Allen}, L.~E., {et~al.} 2004, \apjs, 154, 10

\bibitem[{{Fern{\'a}ndez} \& {Comer{\'o}n}(2005)}]{Fernandez2005}
{Fern{\'a}ndez}, M. \& {Comer{\'o}n}, F. 2005, \aap, 440, 1119

\bibitem[{{Fitzpatrick}(1999)}]{Fitzpatrick1999}
{Fitzpatrick}, E.~L. 1999, \pasp, 111, 63

\bibitem[{{Flaccomio} {et~al.}(2009){Flaccomio}, {Stelzer}, {Sciortino},
  {Micela}, {Pillitteri}, \& {Testi}}]{Flaccomio2009}
{Flaccomio}, E., {Stelzer}, B., {Sciortino}, S., {et~al.} 2009, \aap, 505, 695

\bibitem[{{Flaherty} {et~al.}(2007){Flaherty}, {Pipher}, {Megeath}, {Winston},
  {Gutermuth}, {Muzerolle}, {Allen}, \& {Fazio}}]{Flaherty2007}
{Flaherty}, K.~M., {Pipher}, J.~L., {Megeath}, S.~T., {et~al.} 2007, \apj, 663,
  1069

\bibitem[{{Gagn{\'e}} {et~al.}(2004){Gagn{\'e}}, {Skinner}, \&
  {Daniel}}]{Gagne2004}
{Gagn{\'e}}, M., {Skinner}, S.~L., \& {Daniel}, K.~J. 2004, \apj, 613, 393

\bibitem[{{Giardino} {et~al.}(2007){Giardino}, {Favata}, {Pillitteri},
  {Flaccomio}, {Micela}, \& {Sciortino}}]{Giardino2007}
{Giardino}, G., {Favata}, F., {Pillitteri}, I., {et~al.} 2007, \aap, 475, 891

\bibitem[{{Greene} \& {Meyer}(1995)}]{Greene1995}
{Greene}, T.~P. \& {Meyer}, M.~R. 1995, \apj, 450, 233

\bibitem[{{Greene} {et~al.}(1994){Greene}, {Wilking}, {Andre}, {Young}, \&
  {Lada}}]{Greene1994}
{Greene}, T.~P., {Wilking}, B.~A., {Andre}, P., {Young}, E.~T., \& {Lada},
  C.~J. 1994, \apj, 434, 614

\bibitem[{{Greene} \& {Young}(1992)}]{Greene1992}
{Greene}, T.~P. \& {Young}, E.~T. 1992, \apj, 395, 516

\bibitem[{{Grosso} {et~al.}(2003){Grosso}, {Alves}, {Wood}, {Neuh{\"a}user},
  {Montmerle}, \& {Bjorkman}}]{Grosso2003}
{Grosso}, N., {Alves}, J., {Wood}, K., {et~al.} 2003, \apj, 586, 296

\bibitem[{{Guieu} {et~al.}(2009){Guieu}, {Rebull}, {Stauffer}, {Hillenbrand},
  {Carpenter}, {Noriega-Crespo}, {Padgett}, {Cole}, {Carey}, {Stapelfeldt}, \&
  {Strom}}]{Guieu2009}
{Guieu}, S., {Rebull}, L.~M., {Stauffer}, J.~R., {et~al.} 2009, \apj, 697, 787

\bibitem[{{Gutermuth} {et~al.}(2009){Gutermuth}, {Megeath}, {Myers}, {Allen},
  {Pipher}, \& {Fazio}}]{Gutermuth2009}
{Gutermuth}, R.~A., {Megeath}, S.~T., {Myers}, P.~C., {et~al.} 2009, \apjs,
  184, 18

\bibitem[{{Haisch} {et~al.}(2001){Haisch}, {Lada}, \& {Lada}}]{Haisch2001}
{Haisch}, Jr., K.~E., {Lada}, E.~A., \& {Lada}, C.~J. 2001, \apjl, 553, L153

\bibitem[{{Hennebelle} \& {Chabrier}(2008)}]{Hennebelle2008}
{Hennebelle}, P. \& {Chabrier}, G. 2008, \apj, 684, 395

\bibitem[{{Imanishi} {et~al.}(2001){Imanishi}, {Tsujimoto}, \&
  {Koyama}}]{Imanishi2001}
{Imanishi}, K., {Tsujimoto}, M., \& {Koyama}, K. 2001, \apj, 563, 361

\bibitem[{{Jansen} {et~al.}(2001){Jansen}, {Lumb}, {Altieri}, {Clavel}, {Ehle},
  {Erd}, {Gabriel}, {Guainazzi}, {Gondoin}, {Much}, {Munoz}, {Santos},
  {Schartel}, {Texier}, \& {Vacanti}}]{Jansen2001}
{Jansen}, F., {Lumb}, D., {Altieri}, B., {et~al.} 2001, \aap, 365, L1

\bibitem[{{Kenyon} \& {Hartmann}(1995)}]{Kenyon1995}
{Kenyon}, S.~J. \& {Hartmann}, L. 1995, \apjs, 101, 117

\bibitem[{{Kirkpatrick} {et~al.}(2006){Kirkpatrick}, {Barman}, {Burgasser},
  {McGovern}, {McLean}, {Tinney}, \& {Lowrance}}]{Kirkpatrick2006}
{Kirkpatrick}, J.~D., {Barman}, T.~S., {Burgasser}, A.~J., {et~al.} 2006, \apj,
  639, 1120

\bibitem[{{Larson}(1973)}]{Larson1973}
{Larson}, R.~B. 1973, \mnras, 161, 133

\bibitem[{{Lodieu} {et~al.}(2009){Lodieu}, {Zapatero Osorio}, {Rebolo},
  {Mart{\'{\i}}n}, \& {Hambly}}]{Lodieu2009}
{Lodieu}, N., {Zapatero Osorio}, M.~R., {Rebolo}, R., {Mart{\'{\i}}n}, E.~L.,
  \& {Hambly}, N.~C. 2009, \aap, 505, 1115

\bibitem[{{Lombardi} \& {Alves}(2001)}]{Lombardi2001}
{Lombardi}, M. \& {Alves}, J. 2001, \aap, 377, 1023

\bibitem[{{Lombardi} {et~al.}(2008){Lombardi}, {Lada}, \&
  {Alves}}]{Lombardi2008}
{Lombardi}, M., {Lada}, C.~J., \& {Alves}, J. 2008, \aap, 489, 143

\bibitem[{{Lucas} \& {Roche}(2000)}]{Lucas2000}
{Lucas}, P.~W. \& {Roche}, P.~F. 2000, \mnras, 314, 858

\bibitem[{{Lucas} {et~al.}(2001){Lucas}, {Roche}, {Allard}, \&
  {Hauschildt}}]{Lucas2001}
{Lucas}, P.~W., {Roche}, P.~F., {Allard}, F., \& {Hauschildt}, P.~H. 2001,
  \mnras, 326, 695

\bibitem[{{Luhman}(2004)}]{Luhman2004}
{Luhman}, K.~L. 2004, \apj, 617, 1216

\bibitem[{{Luhman}(2007)}]{Luhman2007}
{Luhman}, K.~L. 2007, \apjs, 173, 104

\bibitem[{{Luhman} {et~al.}(2010){Luhman}, {Allen}, {Espaillat}, {Hartmann}, \&
  {Calvet}}]{Luhman2010}
{Luhman}, K.~L., {Allen}, P.~R., {Espaillat}, C., {Hartmann}, L., \& {Calvet},
  N. 2010, \apjs, 186, 111

\bibitem[{{Luhman} {et~al.}(2007{\natexlab{a}}){Luhman}, {Allers}, {Jaffe},
  {Cushing}, {Williams}, {Slesnick}, \& {Vacca}}]{Luhman2007a}
{Luhman}, K.~L., {Allers}, K.~N., {Jaffe}, D.~T., {et~al.} 2007{\natexlab{a}},
  \apj, 659, 1629

\bibitem[{{Luhman} {et~al.}(2003{\natexlab{a}}){Luhman}, {Brice{\~n}o},
  {Stauffer}, {Hartmann}, {Barrado y Navascu{\'e}s}, \&
  {Caldwell}}]{Luhman2003b}
{Luhman}, K.~L., {Brice{\~n}o}, C., {Stauffer}, J.~R., {et~al.}
  2003{\natexlab{a}}, \apj, 590, 348

\bibitem[{{Luhman} {et~al.}(2007{\natexlab{b}}){Luhman}, {Joergens}, {Lada},
  {Muzerolle}, {Pascucci}, \& {White}}]{Luhman2007c}
{Luhman}, K.~L., {Joergens}, V., {Lada}, C., {et~al.} 2007{\natexlab{b}},
  Protostars and Planets V, 443

\bibitem[{{Luhman} {et~al.}(1997){Luhman}, {Liebert}, \& {Rieke}}]{Luhman1997}
{Luhman}, K.~L., {Liebert}, J., \& {Rieke}, G.~H. 1997, \apjl, 489, L165+

\bibitem[{{Luhman} {et~al.}(2009){Luhman}, {Mamajek}, {Allen}, \&
  {Cruz}}]{Luhman2009}
{Luhman}, K.~L., {Mamajek}, E.~E., {Allen}, P.~R., \& {Cruz}, K.~L. 2009, \apj,
  703, 399

\bibitem[{{Luhman} \& {Rieke}(1999)}]{Luhman1999}
{Luhman}, K.~L. \& {Rieke}, G.~H. 1999, \apj, 525, 440

\bibitem[{{Luhman} {et~al.}(2003{\natexlab{b}}){Luhman}, {Stauffer}, {Muench},
  {Rieke}, {Lada}, {Bouvier}, \& {Lada}}]{Luhman2003}
{Luhman}, K.~L., {Stauffer}, J.~R., {Muench}, A.~A., {et~al.}
  2003{\natexlab{b}}, \apj, 593, 1093

\bibitem[{{Mamajek}(2008)}]{Mamajek2008}
{Mamajek}, E.~E. 2008, Astronomische Nachrichten, 329, 10

\bibitem[{{Marmo}(2007)}]{Marmo2007}
{Marmo}, C. 2007, in Astronomical Society of the Pacific Conference Series,
  Vol. 376, Astronomical Data Analysis Software and Systems XVI, ed.
  {R.~A.~Shaw, F.~Hill, \& D.~J.~Bell}, 285--+

\bibitem[{{Marsh} {et~al.}(2009){Marsh}, {Kirkpatrick}, \&
  {Plavchan}}]{Marsh2009}
{Marsh}, K.~A., {Kirkpatrick}, J.~D., \& {Plavchan}, P. 2009, ArXiv e-prints

\bibitem[{{Martin} {et~al.}(1998){Martin}, {Montmerle}, {Gregorio-Hetem}, \&
  {Casanova}}]{Martin1998}
{Martin}, E.~L., {Montmerle}, T., {Gregorio-Hetem}, J., \& {Casanova}, S. 1998,
  \mnras, 300, 733

\bibitem[{{Megeath} {et~al.}(2004){Megeath}, {Allen}, {Gutermuth}, {Pipher},
  {Myers}, {Calvet}, {Hartmann}, {Muzerolle}, \& {Fazio}}]{Megeath2004}
{Megeath}, S.~T., {Allen}, L.~E., {Gutermuth}, R.~A., {et~al.} 2004, \apjs,
  154, 367

\bibitem[{{Miyazaki} {et~al.}(2002){Miyazaki}, {Komiyama}, {Sekiguchi},
  {Okamura}, {Doi}, {Furusawa}, {Hamabe}, {Imi}, {Kimura}, {Nakata}, {Okada},
  {Ouchi}, {Shimasaku}, {Yagi}, \& {Yasuda}}]{Miyazaki2002}
{Miyazaki}, S., {Komiyama}, Y., {Sekiguchi}, M., {et~al.} 2002, \pasj, 54, 833

\bibitem[{{Moraux} {et~al.}(2007){Moraux}, {Bouvier}, {Stauffer}, {Barrado y
  Navascu{\'e}s}, \& {Cuillandre}}]{Moraux2007}
{Moraux}, E., {Bouvier}, J., {Stauffer}, J.~R., {Barrado y Navascu{\'e}s}, D.,
  \& {Cuillandre}, J. 2007, \aap, 471, 499

\bibitem[{{Natta} {et~al.}(2002){Natta}, {Testi}, {Comer{\'o}n}, {Oliva},
  {D'Antona}, {Baffa}, {Comoretto}, \& {Gennari}}]{Natta2002}
{Natta}, A., {Testi}, L., {Comer{\'o}n}, F., {et~al.} 2002, \aap, 393, 597

\bibitem[{{Natta} {et~al.}(2006){Natta}, {Testi}, \& {Randich}}]{Natta2006}
{Natta}, A., {Testi}, L., \& {Randich}, S. 2006, \aap, 452, 245

\bibitem[{{Ouchi} {et~al.}(2004){Ouchi}, {Shimasaku}, {Okamura}, {Furusawa},
  {Kashikawa}, {Ota}, {Doi}, {Hamabe}, {Kimura}, {Komiyama}, {Miyazaki},
  {Miyazaki}, {Nakata}, {Sekiguchi}, {Yagi}, \& {Yasuda}}]{Ouchi2004}
{Ouchi}, M., {Shimasaku}, K., {Okamura}, S., {et~al.} 2004, \apj, 611, 660

\bibitem[{{Padgett} {et~al.}(2008){Padgett}, {Rebull}, {Stapelfeldt},
  {Chapman}, {Lai}, {Mundy}, {Evans}, {Brooke}, {Cieza}, {Spiesman},
  {Noriega-Crespo}, {McCabe}, {Allen}, {Blake}, {Harvey}, {Huard},
  {J{\o}rgensen}, {Koerner}, {Myers}, {Sargent}, {Teuben}, {van Dishoeck},
  {Wahhaj}, \& {Young}}]{Padgett2008}
{Padgett}, D.~L., {Rebull}, L.~M., {Stapelfeldt}, K.~R., {et~al.} 2008, \apj,
  672, 1013

\bibitem[{{Padoan} {et~al.}(2007){Padoan}, {Nordlund}, {Kritsuk}, {Norman}, \&
  {Li}}]{Padoan2007}
{Padoan}, P., {Nordlund}, {\AA}., {Kritsuk}, A.~G., {Norman}, M.~L., \& {Li},
  P.~S. 2007, \apj, 661, 972

\bibitem[{{Pickles}(1998)}]{Pickles1998}
{Pickles}, A.~J. 1998, \pasp, 110, 863

\bibitem[{{Preibisch} \& {Zinnecker}(1999)}]{Preibisch1999}
{Preibisch}, T. \& {Zinnecker}, H. 1999, \aj, 117, 2381

\bibitem[{{Puget} {et~al.}(2004){Puget}, {Stadler}, {Doyon}, {Gigan},
  {Thibault}, {Luppino}, {Barrick}, {Benedict}, {Forveille}, {Rambold},
  {Thomas}, {Vermeulen}, {Ward}, {Beuzit}, {Feautrier}, {Magnard}, {Mella},
  {Preis}, {Vallee}, {Wang}, {Lin}, {Hall}, \& {Hodapp}}]{Puget2004}
{Puget}, P., {Stadler}, E., {Doyon}, R., {et~al.} 2004, in Society of
  Photo-Optical Instrumentation Engineers (SPIE) Conference Series, Vol. 5492,
  Society of Photo-Optical Instrumentation Engineers (SPIE) Conference Series,
  ed. {A.~F.~M.~Moorwood \& M.~Iye}, 978--987

\bibitem[{{Rayner} {et~al.}(2003){Rayner}, {Toomey}, {Onaka}, {Denault},
  {Stahlberger}, {Vacca}, {Cushing}, \& {Wang}}]{Rayner2003}
{Rayner}, J.~T., {Toomey}, D.~W., {Onaka}, P.~M., {et~al.} 2003, \pasp, 115,
  362

\bibitem[{{Reipurth} \& {Clarke}(2001)}]{Reipurth2001}
{Reipurth}, B. \& {Clarke}, C. 2001, \aj, 122, 432

\bibitem[{{Ridge} {et~al.}(2006){Ridge}, {Di Francesco}, {Kirk}, {Li},
  {Goodman}, {Alves}, {Arce}, {Borkin}, {Caselli}, {Foster}, {Heyer},
  {Johnstone}, {Kosslyn}, {Lombardi}, {Pineda}, {Schnee}, \&
  {Tafalla}}]{Ridge2006}
{Ridge}, N.~A., {Di Francesco}, J., {Kirk}, H., {et~al.} 2006, \aj, 131, 2921

\bibitem[{{Rieke} \& {Lebofsky}(1985)}]{Rieke1985}
{Rieke}, G.~H. \& {Lebofsky}, M.~J. 1985, \apj, 288, 618

\bibitem[{{Rieke} \& {Rieke}(1990)}]{Rieke1990}
{Rieke}, G.~H. \& {Rieke}, M.~J. 1990, \apjl, 362, L21

\bibitem[{{Rieke} {et~al.}(2004){Rieke}, {Young}, {Engelbracht}, {Kelly},
  {Low}, {Haller}, {Beeman}, {Gordon}, {Stansberry}, {Misselt}, {Cadien},
  {Morrison}, {Rivlis}, {Latter}, {Noriega-Crespo}, {Padgett}, {Stapelfeldt},
  {Hines}, {Egami}, {Muzerolle}, {Alonso-Herrero}, {Blaylock}, {Dole}, {Hinz},
  {Le Floc'h}, {Papovich}, {P{\'e}rez-Gonz{\'a}lez}, {Smith}, {Su}, {Bennett},
  {Frayer}, {Henderson}, {Lu}, {Masci}, {Pesenson}, {Rebull}, {Rho}, {Keene},
  {Stolovy}, {Wachter}, {Wheaton}, {Werner}, \& {Richards}}]{Rieke2004}
{Rieke}, G.~H., {Young}, E.~T., {Engelbracht}, C.~W., {et~al.} 2004, \apjs,
  154, 25

\bibitem[{{Robin} {et~al.}(2003){Robin}, {Reyl{\'e}}, {Derri{\`e}re}, \&
  {Picaud}}]{Robin2003}
{Robin}, A.~C., {Reyl{\'e}}, C., {Derri{\`e}re}, S., \& {Picaud}, S. 2003,
  \aap, 409, 523

\bibitem[{{Santiago} {et~al.}(1996){Santiago}, {Gilmore}, \&
  {Elson}}]{Santiago1996}
{Santiago}, B.~X., {Gilmore}, G., \& {Elson}, R.~A.~W. 1996, \mnras, 281, 871

\bibitem[{{Sauter} {et~al.}(2009){Sauter}, {Wolf}, {Launhardt}, {Padgett},
  {Stapelfeldt}, {Pinte}, {Duch{\^e}ne}, {M{\'e}nard}, {McCabe}, {Pontoppidan},
  {Dunham}, {Bourke}, \& {Chen}}]{Sauter2009}
{Sauter}, J., {Wolf}, S., {Launhardt}, R., {et~al.} 2009, \aap, 505, 1167

\bibitem[{{Schmidt}(2002)}]{Schmidt2002}
{Schmidt}, E.~G. 2002, \aj, 123, 965

\bibitem[{{Scholz} {et~al.}(2009){Scholz}, {Geers}, {Jayawardhana}, {Fissel},
  {Lee}, {Lafreniere}, \& {Tamura}}]{Scholz2009}
{Scholz}, A., {Geers}, V., {Jayawardhana}, R., {et~al.} 2009, \apj, 702, 805

\bibitem[{{Sciortino} {et~al.}(2006){Sciortino}, {Pillitteri}, {Damiani},
  {Flaccomio}, {Micela}, {Stelzer}, {Favata}, {Giardino}, {Grosso},
  {Montmerle}, {Palla}, \& {Testi}}]{Sciortino2006}
{Sciortino}, S., {Pillitteri}, I., {Damiani}, F., {et~al.} 2006, in ESA Special
  Publication, Vol. 604, The X-ray Universe 2005, ed. {A.~Wilson}, 111--+

\bibitem[{{Shu} {et~al.}(1987){Shu}, {Adams}, \& {Lizano}}]{Shu1987}
{Shu}, F.~H., {Adams}, F.~C., \& {Lizano}, S. 1987, \araa, 25, 23

\bibitem[{{Stamatellos} \& {Whitworth}(2009)}]{Stamatellos2009}
{Stamatellos}, D. \& {Whitworth}, A.~P. 2009, \mnras, 392, 413

\bibitem[{{Stern} {et~al.}(2005){Stern}, {Eisenhardt}, {Gorjian}, {Kochanek},
  {Caldwell}, {Eisenstein}, {Brodwin}, {Brown}, {Cool}, {Dey}, {Green},
  {Jannuzi}, {Murray}, {Pahre}, \& {Willner}}]{Stern2005}
{Stern}, D., {Eisenhardt}, P., {Gorjian}, V., {et~al.} 2005, \apj, 631, 163

\bibitem[{{Strom} {et~al.}(1995){Strom}, {Kepner}, \& {Strom}}]{Strom1995}
{Strom}, K.~M., {Kepner}, J., \& {Strom}, S.~E. 1995, \apj, 438, 813

\bibitem[{{Str{\"u}der} {et~al.}(2001){Str{\"u}der}, {Briel}, {Dennerl},
  {Hartmann}, {Kendziorra}, {Meidinger}, {Pfeffermann}, {Reppin}, {Aschenbach},
  {Bornemann}, {Br{\"a}uninger}, {Burkert}, {Elender}, {Freyberg}, {Haberl},
  {Hartner}, {Heuschmann}, {Hippmann}, {Kastelic}, {Kemmer}, {Kettenring},
  {Kink}, {Krause}, {M{\"u}ller}, {Oppitz}, {Pietsch}, {Popp}, {Predehl},
  {Read}, {Stephan}, {St{\"o}tter}, {Tr{\"u}mper}, {Holl}, {Kemmer}, {Soltau},
  {St{\"o}tter}, {Weber}, {Weichert}, {von Zanthier}, {Carathanassis}, {Lutz},
  {Richter}, {Solc}, {B{\"o}ttcher}, {Kuster}, {Staubert}, {Abbey}, {Holland},
  {Turner}, {Balasini}, {Bignami}, {La Palombara}, {Villa}, {Buttler},
  {Gianini}, {Lain{\'e}}, {Lumb}, \& {Dhez}}]{Struder2001}
{Str{\"u}der}, L., {Briel}, U., {Dennerl}, K., {et~al.} 2001, \aap, 365, L18

\bibitem[{{Turner} {et~al.}(2001){Turner}, {Abbey}, {Arnaud}, {Balasini},
  {Barbera}, {Belsole}, {Bennie}, {Bernard}, {Bignami}, {Boer}, {Briel},
  {Butler}, {Cara}, {Chabaud}, {Cole}, {Collura}, {Conte}, {Cros}, {Denby},
  {Dhez}, {Di Coco}, {Dowson}, {Ferrando}, {Ghizzardi}, {Gianotti}, {Goodall},
  {Gretton}, {Griffiths}, {Hainaut}, {Hochedez}, {Holland}, {Jourdain},
  {Kendziorra}, {Lagostina}, {Laine}, {La Palombara}, {Lortholary}, {Lumb},
  {Marty}, {Molendi}, {Pigot}, {Poindron}, {Pounds}, {Reeves}, {Reppin},
  {Rothenflug}, {Salvetat}, {Sauvageot}, {Schmitt}, {Sembay}, {Short},
  {Spragg}, {Stephen}, {Str{\"u}der}, {Tiengo}, {Trifoglio}, {Tr{\"u}mper},
  {Vercellone}, {Vigroux}, {Villa}, {Ward}, {Whitehead}, \&
  {Zonca}}]{Turner2001}
{Turner}, M.~J.~L., {Abbey}, A., {Arnaud}, M., {et~al.} 2001, \aap, 365, L27

\bibitem[{{Wainscoat} {et~al.}(1992){Wainscoat}, {Cohen}, {Volk}, {Walker}, \&
  {Schwartz}}]{Wainscoat1992}
{Wainscoat}, R.~J., {Cohen}, M., {Volk}, K., {Walker}, H.~J., \& {Schwartz},
  D.~E. 1992, \apjs, 83, 111

\bibitem[{{Weights} {et~al.}(2009){Weights}, {Lucas}, {Roche}, {Pinfield}, \&
  {Riddick}}]{Weights2009}
{Weights}, D.~J., {Lucas}, P.~W., {Roche}, P.~F., {Pinfield}, D.~J., \&
  {Riddick}, F. 2009, \mnras, 392, 817

\bibitem[{{Whitworth} \& {Goodwin}(2005)}]{Whitworth2005}
{Whitworth}, A.~P. \& {Goodwin}, S.~P. 2005, Astronomische Nachrichten, 326,
  899

\bibitem[{{Wilking} {et~al.}(2008){Wilking}, {Gagn{\'e}}, \&
  {Allen}}]{Wilking2008}
{Wilking}, B.~A., {Gagn{\'e}}, M., \& {Allen}, L.~E. 2008, {Star Formation in
  the {$\rho$} Ophiuchi Molecular Cloud}, ed. B.~Reipurth, 351--+

\bibitem[{{Wilking} {et~al.}(1999){Wilking}, {Greene}, \&
  {Meyer}}]{Wilking1999}
{Wilking}, B.~A., {Greene}, T.~P., \& {Meyer}, M.~R. 1999, \aj, 117, 469

\bibitem[{{Wilking} \& {Lada}(1983)}]{Wilking1983}
{Wilking}, B.~A. \& {Lada}, C.~J. 1983, \apj, 274, 698

\bibitem[{{Wilking} {et~al.}(1989){Wilking}, {Lada}, \& {Young}}]{Wilking1989}
{Wilking}, B.~A., {Lada}, C.~J., \& {Young}, E.~T. 1989, \apj, 340, 823

\bibitem[{{Wilking} {et~al.}(2005){Wilking}, {Meyer}, {Robinson}, \&
  {Greene}}]{Wilking2005}
{Wilking}, B.~A., {Meyer}, M.~R., {Robinson}, J.~G., \& {Greene}, T.~P. 2005,
  \aj, 130, 1733

\bibitem[{{Yagi} {et~al.}(2002){Yagi}, {Kashikawa}, {Sekiguchi}, {Doi},
  {Yasuda}, {Shimasaku}, \& {Okamura}}]{Yagi2002}
{Yagi}, M., {Kashikawa}, N., {Sekiguchi}, M., {et~al.} 2002, \aj, 123, 66

\bibitem[{{Zapatero Osorio} {et~al.}(2002){Zapatero Osorio}, {B{\'e}jar},
  {Mart{\'{\i}}n}, {Rebolo}, {Barrado y Navascu{\'e}s}, {Mundt},
  {Eisl{\"o}ffel}, \& {Caballero}}]{ZapateroOsorio2002}
{Zapatero Osorio}, M.~R., {B{\'e}jar}, V.~J.~S., {Mart{\'{\i}}n}, E.~L.,
  {et~al.} 2002, \apj, 578, 536

\end{thebibliography}


\onllongtab{4}{
\begin{landscape}
\tiny
\begin{longtable}{@{ }l@{ }l@{ }c@{ }c@{ }c@{ }c@{ }c@{ }c@{ }c@{ }c@{ }c@{ }c@{ }c@{ }c@{ }c@{ }c@{ }c@{ }c@{ }}
\caption{\label{stars} Candidate Members of the $\rho$~Ophiuchi Molecular Cloud}\\
\hline\hline
\multicolumn{2}{l}{CFHTWIR-Oph\footnotemark[1]} & \emph{$i'$}\footnotemark[2] & \emph{$z'$}\footnotemark[2] &  $\textit{J}$\footnotemark[3]     & $\textit{H}$\footnotemark[3]   &  $\textit{K$_{s}$}$\footnotemark[3]  &  
$[3.6]$\footnotemark[4] & $[4.8]$\footnotemark[4] & $[5.4]$\footnotemark[4] & $[8.0]$\footnotemark[4] & $[24]$\footnotemark[4] & SpT\footnotemark[5] & A$_{\emph{V}}$\footnotemark[5] & Literature\footnotemark[6] &  Flag\footnotemark[7]  & Other name \\
~ & ~ & (mag) &  (mag) & (mag) & (mag) &  (mag)  &  (mag) & (mag) & (mag) & (mag) & (mag) & ~ & (mag) & ~ & ~ & ~ \\

\hline
\endfirsthead
\caption{continued.}\\
\hline\hline
\multicolumn{2}{l}{CFHTWIR-Oph} & \emph{$i'$} & \emph{$z'$} &  $\textit{J}$  & $\textit{H}$ &  $\textit{K$_{s}$}$ &  
$[3.6]$ & $[4.8]$ & $[5.4]$ & $[8.0]$  & $[24]$  & SpT  & A$_{\emph{V}}$  & Literature  &  Flag & Other name \\
~ & ~ & (mag) &  (mag) & (mag) & (mag) &  (mag)  &  (mag) & (mag) & (mag) & (mag) & (mag) & ~ & (mag) & ~ & ~& ~  \\
\hline
\endhead
\hline
\endfoot

1   & \object{J162519.15-241927.4} &     &    & 19.07$\pm$0.05 & 17.33$\pm$0.05 & 15.86$\pm$0.05 & 13.37$\pm$0.06 & 12.35$\pm$0.06 & 11.56$\pm$0.07 & 10.57$\pm$0.08 & 7.3 $\pm$0.29  &  &  &  & ex1,ex2 & \\ 
2   & \object{J162522.61-243453.5} &     &    & 17.69$\pm$0.05 & 14.87$\pm$0.05 & 12.72$\pm$0.05 & 9.96 $\pm$0.06 & 9.12 $\pm$0.06 & 8.57 $\pm$0.05 & 8.14 $\pm$0.05 & 5.9 $\pm$0.11  &  &  &  & ex1,ex2,R & \\ 
3   & \object{J162523.78-242258.4} &     &    & 16.3 $\pm$0.05 & 14.85$\pm$0.05 & 13.97$\pm$0.05 & 13.17$\pm$0.06 & 13.05$\pm$0.06 &    &    &     &  &  &  &  & \\ 
4   & \object{J162532.41-243405.2} &     &    & 14.88$\pm$0.05 & 13.93$\pm$0.05 & 13.34$\pm$0.05 & 12.71$\pm$0.06 & 12.53$\pm$0.06 & 12.37$\pm$0.1  &    &     & M6.50$^{+0.25}_{-0.25}$ &     2.5$^{+0.1}_{-0.1}$   & & & \\ 
5   & \object{J162541.28-243014.8} &     &    & 21.92$\pm$0.22 & 19.08$\pm$0.08 & 17.25$\pm$0.05 &    &    &    &    &     &  &  &  &  & \\ 
6   & \object{J162546.63-242336.4} &     &    & 18.87$\pm$0.05 & 16.69$\pm$0.05 & 15.12$\pm$0.05 & 12.58$\pm$0.06 & 11.86$\pm$0.06 & 11.52$\pm$0.08 & 11.01$\pm$0.25 & 6.73$\pm$0.14  &  &  &  & ex1,ex2 & \\ 
7   & \object{J162546.98-241154.1} &     &    & 18.34$\pm$0.05 & 16.85$\pm$0.05 & 15.68$\pm$0.05 & 14.66$\pm$0.07 & 14.36$\pm$0.07 &    &    & 8.71$\pm$0.26  &  &  &  & ex2 & \\ 
8   & \object{J162557.69-242318.4} &     & 23.11$\pm$0.04 & 17.68$\pm$0.05 & 14.84$\pm$0.05 & 13.11$\pm$0.05 & 11.81$\pm$0.06 & 11.35$\pm$0.05 & 11.19$\pm$0.07 &    &     &  &  &  &  & \\ 
9   & \object{J162603.28-243025.9} &     & 21.38$\pm$0.03 & 17.76$\pm$0.05 & 16.33$\pm$0.05 & 15.32$\pm$0.05 & 14.12$\pm$0.07 & 13.63$\pm$0.06 &    &    &     &  &  &  &  & \\ 
10  & \object{J162607.25-242116.9} &     &    & 21.74$\pm$0.15 & 18.6 $\pm$0.06 & 16.69$\pm$0.05 & 14.52$\pm$0.09 & 14.04$\pm$0.07 &    &    &     &  &  &  &  & \\ 
11  & \object{J162607.96-241723.0} &     &    & 22.26$\pm$0.2  & 19.25$\pm$0.07 & 17.31$\pm$0.05 & 15.65$\pm$0.08 & 15.35$\pm$0.09 &    &    &     &  &  &  &  & \\ 
12  & \object{J162609.99-241440.1} &     & 21.7 $\pm$0.03 & 18.15$\pm$0.05 & 16.35$\pm$0.05 & 15.25$\pm$0.05 & 14.55$\pm$0.06 & 14.29$\pm$0.07 &    &    &     &  &  &  &  & \\ 
13  & \object{J162611.74-242431.0} &     &    & 21.2 $\pm$0.1  & 18.45$\pm$0.06 & 16.13$\pm$0.05 & 14.3 $\pm$0.1  & 13.83$\pm$0.08 &    &    &     &  &  &  &  & \\ 
14  & \object{J162613.20-241910.5} &     &    & 22.17$\pm$0.19 & 18.77$\pm$0.06 & 16.4 $\pm$0.05 & 14.54$\pm$0.06 & 14.12$\pm$0.06 & 13.59$\pm$0.17 &    &     &  &  &  &  & \\ 
15  & \object{J162616.33-243930.7} &  21.52$\pm$0.01 & 19.2 $\pm$0.03 & 15.79$\pm$0.05 & 14.15$\pm$0.05 & 13.11$\pm$0.05 & 12.42$\pm$0.06 & 12.19$\pm$0.06 & 12.02$\pm$0.08 & 12.19$\pm$0.29 &     &  &  & 8 &  & \\ 
16  & \object{J162618.58-242951.6} &     & 21.9 $\pm$0.03 & 17.09$\pm$0.05 & 14.92$\pm$0.05 & 13.55$\pm$0.05 & 12.14$\pm$0.06 & 11.61$\pm$0.06 & 11.0 $\pm$0.07 & 10.18$\pm$0.14 & 6.99$\pm$0.12  &  &  & 8 & var,ex1,ex2 & \\ 
17  & \object{J162619.21-244130.1} &  22.61$\pm$0.02 & 20.86$\pm$0.03 & 17.44$\pm$0.05 & 15.8 $\pm$0.05 & 14.61$\pm$0.05 & 13.98$\pm$0.06 & 13.83$\pm$0.06 & 13.92$\pm$0.18 &    &     &  &  &  &  & \\ 
18  & \object{J162619.41-242743.9} &     & 22.92$\pm$0.04 & 18.92$\pm$0.05 & 17.16$\pm$0.05 & 16.04$\pm$0.05 &    &    &    &    &     &  &  &  &  & \\ 
19  & \object{J162622.27-243709.5} &     & 22.36$\pm$0.03 & 19.15$\pm$0.05 & 17.1 $\pm$0.05 & 15.86$\pm$0.05 & 15.01$\pm$0.08 & 14.81$\pm$0.09 &    &    &     &  &  &  &  & \\ 
20  & \object{J162622.89-235834.2} &     &    & 18.01$\pm$0.05 & 16.73$\pm$0.05 & 15.8 $\pm$0.05 &    &    &    &    &     &  &  &  &  & \\ 
21  & \object{J162624.29-241549.8} &     & 23.13$\pm$0.04 & 18.78$\pm$0.05 & 15.85$\pm$0.05 & 13.83$\pm$0.05 & 11.7 $\pm$0.06 & 10.96$\pm$0.06 &    & 9.64 $\pm$0.1  & 6.88$\pm$0.16  &  &  &  & var,ex2 & AOC J162624.29-241549.7 \\ 
22  & \object{J162625.10-244132.7} &  21.78$\pm$0.01 & 20.05$\pm$0.03 & 17.03$\pm$0.05 & 15.52$\pm$0.05 & 14.54$\pm$0.05 & 14.06$\pm$0.06 & 13.9 $\pm$0.06 & 13.83$\pm$0.17 &    &     &  &  &  &  & \\ 
23  & \object{J162625.98-243313.8} &     &    & 22.2 $\pm$0.18 & 19.39$\pm$0.07 & 17.11$\pm$0.05 & 15.2 $\pm$0.08 & 14.83$\pm$0.09 &    &    &     &  &  &  &  & \\ 
24  & \object{J162626.41-243306.4} &     &    & 22.13$\pm$0.19 & 19.45$\pm$0.08 & 17.45$\pm$0.05 & 15.73$\pm$0.11 & 15.28$\pm$0.12 &    &    &     &  &  &  &  & \\ 
25  & \object{J162633.80-241854.0} &     &    & 22.42$\pm$0.22 & 19.31$\pm$0.07 & 17.39$\pm$0.05 &    &    &    &    &     &  &  &  &  & \\ 
26  & \object{J162634.01-243455.6} &     &    & 22.45$\pm$0.25 & 19.31$\pm$0.07 & 17.38$\pm$0.05 & 15.76$\pm$0.14 & 15.29$\pm$0.1  &    &    &     &  &  &  &  & \\ 
27  & \object{J162634.94-243012.1} &     &    & 22.32$\pm$0.26 & 18.35$\pm$0.06 & 15.94$\pm$0.05 & 14.12$\pm$0.06 & 13.76$\pm$0.06 & 13.41$\pm$0.17 &    &     &  &  &  &  & \\ 
28  & \object{J162635.28-244239.2} &     & 22.55$\pm$0.03 & 19.68$\pm$0.06 & 17.79$\pm$0.05 & 16.36$\pm$0.05 & 13.98$\pm$0.06 & 12.97$\pm$0.06 & 11.99$\pm$0.06 & 10.89$\pm$0.06 &     &  &  &  & ex1 & \\ 
29  & \object{J162635.96-242058.9} &     &    & 18.84$\pm$0.05 & 15.65$\pm$0.05 & 13.46$\pm$0.05 & 11.72$\pm$0.06 & 10.83$\pm$0.05 & 10.25$\pm$0.06 & 9.69 $\pm$0.1  &     &  &  &  & var,ex1 & \\ 
30  & \object{J162636.82-241900.3} &  20.42$\pm$0.01  & 18.85$\pm$0.03 & 16.24$\pm$0.05 & 14.35$\pm$0.05 & 13.06$\pm$0.05 & 13.52$\pm$0.06 & 12.36$\pm$0.06 & 11.23$\pm$0.08 & 10.66$\pm$0.07 &     &  &  & 1,8 & var,ex1 & 2MASS J16263682-2419002 \\ 
31  & \object{J162637.81-243903.3} &  19.28$\pm$0.01  & 18.52$\pm$0.03 & 14.65$\pm$0.05 & 13.45$\pm$0.05 & 12.66$\pm$0.05 & 12.0 $\pm$0.06 & 11.65$\pm$0.06 & 11.34$\pm$0.06 & 10.75$\pm$0.06 & 7.75$\pm$0.14  &  &  & 3,8 & ex1,ex2 & BKLT J162638-243901\\ 
32  & \object{J162639.67-241803.0} &     &    & 22.55$\pm$0.3  & 19.15$\pm$0.07 & 17.11$\pm$0.05 & 15.62$\pm$0.07 & 15.33$\pm$0.09 &    &    &     &  &  &  &  & \\ 
33  & \object{J162639.69-242206.2} &     & 22.12$\pm$0.03 & 18.16$\pm$0.05 & 16.74$\pm$0.05 & 15.68$\pm$0.05 & 14.35$\pm$0.1  & 14.12$\pm$0.07 &    &    &     &  &  &  &  & \\ 
34  & \object{J162639.92-242233.6} &  22.17$\pm$0.01 & 19.71$\pm$0.03 & 15.97$\pm$0.05 & 14.53$\pm$0.05 & 13.48$\pm$0.05 & 12.21$\pm$0.07 & 11.66$\pm$0.06 & 11.13$\pm$0.24 &    &     &  M8.25$^{+0.5}_{-1.0}$ &      9.70$^{+0.7}_{-0.4}$   & 3,5,8& var & [SKS95] 162338.6-241551\\ 
35  & \object{J162640.10-242807.3} &     &    & 21.45$\pm$0.18 & 18.56$\pm$0.07 & 16.52$\pm$0.05 & 14.89$\pm$0.06 & 14.45$\pm$0.07 & 14.24$\pm$0.29 &    &     &  &  &  &  & \\ 
36  & \object{J162640.65-242427.1} &     &    & 21.6 $\pm$0.19 & 18.37$\pm$0.06 & 15.69$\pm$0.05 & 13.93$\pm$0.07 & 13.44$\pm$0.08 &    &    &     &  &  &  &  & \\ 
37  & \object{J162640.84-243051.1} &     & 22.28$\pm$0.03 & 17.32$\pm$0.05 & 14.77$\pm$0.05 & 13.18$\pm$0.05 & 11.68$\pm$0.06 & 11.13$\pm$0.05 & 10.66$\pm$0.06 & 10.0 $\pm$0.06 & 6.62$\pm$0.11  &  &  & 1,3,4,7,8 & ex1,ex2 & BBRCG 1\\ 
38  & \object{J162641.73-243610.9} &     &    & 22.12$\pm$0.23 & 18.62$\pm$0.06 & 16.28$\pm$0.05 & 14.43$\pm$0.06 & 14.04$\pm$0.06 & 13.58$\pm$0.13 & 13.28$\pm$0.27 &     &  &  &  & ex1,AGN? & \\ 
39  & \object{J162641.87-242343.0} &     &    & 22.59$\pm$0.46 & 19.32$\pm$0.1  & 17.07$\pm$0.05 & 15.05$\pm$0.09 & 14.44$\pm$0.09 &    &    &     &  &  &  &  & \\ 
40  & \object{J162642.74-242427.7} &     &    & 19.44$\pm$0.06 & 15.59$\pm$0.05 & 13.22$\pm$0.05 & 11.65$\pm$0.06 & 11.13$\pm$0.05 & 10.88$\pm$0.06 & 10.66$\pm$0.24 &     &  &  & 3,7,8 & ex1 & BKLT J162642-242429 \\ 
41  & \object{J162643.86-242450.7} &     &    & 21.67$\pm$0.2  & 17.5 $\pm$0.05 & 14.76$\pm$0.05 & 13.0 $\pm$0.06 & 12.45$\pm$0.06 & 12.15$\pm$0.08 &    &     &  &  & 4,8 &  & BKLT J162643-242452\\ 
42  & \object{J162644.51-240408.8} &     & 21.49$\pm$0.03 & 18.64$\pm$0.05 & 17.19$\pm$0.05 & 16.08$\pm$0.05 &    &    &    &    &     &  &  &  &  & \\ 
43  & \object{J162648.40-242835.0} &     &    & 19.31$\pm$0.05 & 15.19$\pm$0.05 & 12.66$\pm$0.05 & 10.23$\pm$0.06 & 9.45 $\pm$0.05 & 8.82 $\pm$0.05 &    & 5.07$\pm$0.13  &  &  & 6,8 & var,ex2 & GDS J162648.3-242834 \\ 
44  & \object{J162648.73-242625.8} &     &    & 19.32$\pm$0.06 & 15.32$\pm$0.05 & 12.92$\pm$0.05 & 11.33$\pm$0.05 & 10.95$\pm$0.06 & 10.6 $\pm$0.06 & 10.58$\pm$0.06 &     &  &  &  &  & \\ 
45  & \object{J162650.88-242607.9} &     &    & 21.73$\pm$0.27 & 19.17$\pm$0.11 & 17.27$\pm$0.05 & 15.81$\pm$0.08 & 15.28$\pm$0.09 &    &    &     &  &  &  &  & \\ 
46  & \object{J162651.97-243039.7} &     &    & 21.3 $\pm$0.14 & 16.43$\pm$0.05 & 13.45$\pm$0.05 & 10.81$\pm$0.06 & 9.84 $\pm$0.06 & 9.25 $\pm$0.05 & 8.53 $\pm$0.05 & 5.6 $\pm$0.1   &  &  & 1,3,8 & var,ex1,ex2 & 2MASS J16265197-2430394\\ 
47  & \object{J162652.26-240146.8} &     & 19.07$\pm$0.03 & 15.94$\pm$0.05 & 14.75$\pm$0.05 & 13.97$\pm$0.05 & 13.35$\pm$0.07 & 13.17$\pm$0.07 & 13.05$\pm$0.11 & 12.96$\pm$0.2  &     & M7.50$^{+0.5}_{-0.5}$ &     5.6$^{+0.3}_{-0.2}$    & & & \\ 
48  & \object{J162652.70-242452.5} &     &    & 21.76$\pm$0.22 & 18.24$\pm$0.06 & 15.8 $\pm$0.05 & 14.07$\pm$0.06 & 13.65$\pm$0.06 & 13.28$\pm$0.1  & 13.33$\pm$0.28 &     &  &  &  &  & \\ 
49  & \object{J162653.46-243236.6} &     &    & 21.83$\pm$0.22 & 16.69$\pm$0.05 & 13.3 $\pm$0.05 & 10.56$\pm$0.06 & 9.73 $\pm$0.05 & 9.29 $\pm$0.05 & 8.99 $\pm$0.05 & 4.89$\pm$0.1   &  &  & 1,3,4,8 & var,ex1,ex2 & 2MASS J16265346-2432362 \\ 
50  & \object{J162654.30-242438.2} &     &    & 21.71$\pm$0.21 & 17.25$\pm$0.05 & 14.04$\pm$0.05 & 10.87$\pm$0.06 & 10.0 $\pm$0.06 & 9.34 $\pm$0.05 & 8.82 $\pm$0.05 & 6.08$\pm$0.1   &  &  & 1,3,4,8 & ex1,ex2 & BKLT J162654-242440\\ 
51  & \object{J162654.77-242702.4} &     & 23.97$\pm$0.07 & 17.92$\pm$0.05 & 14.91$\pm$0.05 & 12.87$\pm$0.05 & 11.13$\pm$0.06 & 10.51$\pm$0.05 & 10.0 $\pm$0.05 & 9.34 $\pm$0.06 & 5.85$\pm$0.1   &  &  & 1,3,4,8 & var,ex1,ex2 & [GY92] 154\\ 
52  & \object{J162655.54-235736.5} &     &    & 16.02$\pm$0.05 & 15.08$\pm$0.05 & 14.39$\pm$0.05 & 14.36$\pm$0.07 & 14.31$\pm$0.08 &    &    &     &  &  &  &  & \\ 
53  & \object{J162656.36-244120.6} &     & 23.23$\pm$0.04 & 18.6 $\pm$0.05 & 16.34$\pm$0.05 & 14.92$\pm$0.05 & 13.61$\pm$0.06 & 13.14$\pm$0.06 & 12.59$\pm$0.08 & 11.94$\pm$0.07 & 9.22$\pm$0.25  &  &  & 4,8 & ex1,ex2 & \\ 
54  & \object{J162658.47-242004.5} &     &    & 20.55$\pm$0.08 & 18.12$\pm$0.06 & 16.68$\pm$0.05 & 15.64$\pm$0.07 & 15.29$\pm$0.11 &    &    &     &  &  &  &  & \\ 
55  & \object{J162658.66-242455.6} &     &    & 20.22$\pm$0.07 & 17.15$\pm$0.05 & 14.84$\pm$0.05 & 12.2 $\pm$0.06 & 11.35$\pm$0.05 & 10.77$\pm$0.06 & 10.1 $\pm$0.06 & 8.03$\pm$0.15  &  &  & 8 & var,ex1,ex2,R & AOC J162658.65-242455.5\\ 
56  & \object{J162703.59-242005.6} &  22.83$\pm$0.02 & 20.41$\pm$0.03 & 17.13$\pm$0.05 & 15.01$\pm$0.05 & 13.72$\pm$0.05 & 12.11$\pm$0.06 & 11.45$\pm$0.05 & 10.83$\pm$0.06 & 10.08$\pm$0.06 & 5.91$\pm$0.1   &  &  & 1,8 & var,ex1,ex2 & ISO-Oph 94\\ 
57  & \object{J162704.02-240246.9} &     &    & 15.06$\pm$0.05 & 13.92$\pm$0.05 & 13.18$\pm$0.05 & 12.53$\pm$0.07 & 12.37$\pm$0.07 & 12.14$\pm$0.08 & 11.96$\pm$0.09 &     &  M7.50$^{+0.5}_{-0.5}$ &     5.6$^{+0.3}_{-0.2}$    & & & \\ 
58  & \object{J162705.94-241840.3} &     & 20.62$\pm$0.03 & 17.05$\pm$0.05 & 15.77$\pm$0.05 & 14.85$\pm$0.05 & 13.75$\pm$0.06 & 13.27$\pm$0.06 &    & 12.03$\pm$0.11 &     &  &  & 4,8 &  & \\ 
59  & \object{J162707.70-243403.5} &     &    & 22.33$\pm$0.33 & 19.0 $\pm$0.08 & 16.58$\pm$0.05 &    &    &    &    &     &  &  &  &  & \\ 
60  & \object{J162709.03-243025.3} &     &    & 22.14$\pm$0.38 & 19.3 $\pm$0.12 & 16.15$\pm$0.05 & 13.6 $\pm$0.06 & 12.95$\pm$0.06 & 12.44$\pm$0.07 & 12.02$\pm$0.13 &     &  &  &  & ex1 & \\ 
61  & \object{J162709.80-243442.0} &     &    & 22.44$\pm$0.31 & 18.84$\pm$0.07 & 16.51$\pm$0.05 & 14.84$\pm$0.06 & 14.41$\pm$0.07 & 14.25$\pm$0.19 &    &     &  &  &  &  & \\ 
62  & \object{J162710.03-242913.4} &  20.71$\pm$0.01  & 19.48$\pm$0.03 & 16.55$\pm$0.05 & 15.14$\pm$0.05 & 14.23$\pm$0.05 & 13.16$\pm$0.06 & 12.73$\pm$0.06 & 12.26$\pm$0.11 &    & 6.85$\pm$0.11  &  M5.50$^{+0.5}_{-1.25}$ &     9.90$^{+1.8}_{-0.9}$  & 2,3,4,8 & var,ex2 & [WGM99] 2408.6-2229\\ 
63  & \object{J162710.17-243545.9} &     &    & 22.01$\pm$0.21 & 18.92$\pm$0.07 & 17.1 $\pm$0.05 &    &    &    &    &     &  &  &  &  & \\ 
64  & \object{J162713.04-243200.4} &     &    & 22.13$\pm$0.39 & 19.28$\pm$0.11 & 17.05$\pm$0.05 & 15.3 $\pm$0.06 & 14.93$\pm$0.07 & 14.33$\pm$0.2  &    &     &  &  &  &  & \\ 
65  & \object{J162713.24-242347.4} &     &    & 20.88$\pm$0.12 & 17.6 $\pm$0.05 & 15.67$\pm$0.05 & 14.18$\pm$0.06 & 13.84$\pm$0.06 & 13.44$\pm$0.14 &    &     &  &  &  &  & \\ 
66  & \object{J162714.34-243132.0} &     & 22.64$\pm$0.03 & 18.42$\pm$0.05 & 16.53$\pm$0.05 & 15.3 $\pm$0.05 & 14.07$\pm$0.06 & 13.49$\pm$0.06 & 13.06$\pm$0.1  & 11.9 $\pm$0.09 & 7.2 $\pm$0.11  &  &  &  & var,ex1,ex2,AGN? & AOC J162714.34-243131.9 \\ 
67  & \object{J162715.69-243845.7} &     &    & 18.76$\pm$0.05 & 15.22$\pm$0.05 & 12.95$\pm$0.05 &    & 9.45 $\pm$0.08 &    & 6.05 $\pm$0.09 &     &  &  & 6\footnotemark[8] &  & WL 20S \\ 
68  & \object{J162715.88-242514.2} &     &    & 20.23$\pm$0.08 & 15.9 $\pm$0.05 & 13.26$\pm$0.05 & 11.03$\pm$0.06 & 10.22$\pm$0.05 & 9.6  $\pm$0.05 & 8.78 $\pm$0.05 & 5.4 $\pm$0.1   &  &  & 1,3,4,7,8 & var,ex1,ex2 & [GY92] 241 \\ 
69  & \object{J162718.53-240722.0} &     &    & 14.76$\pm$0.05 & 13.55$\pm$0.05 & 12.79$\pm$0.05 & 12.3 $\pm$0.07 & 12.04$\pm$0.06 & 11.86$\pm$0.07 & 11.84$\pm$0.1  &     &  &  &  &  & \\ 
70  & \object{J162719.39-242600.4} &     &    & 21.71$\pm$0.34 & 18.71$\pm$0.08 & 16.79$\pm$0.05 & 15.32$\pm$0.08 & 14.87$\pm$0.08 &    &    &     &  &  &  &  & [AMD2002] J162719-242601 \\ 
71  & \object{J162719.45-242049.0} &     &    & 21.73$\pm$0.21 & 18.35$\pm$0.06 & 16.17$\pm$0.05 & 14.68$\pm$0.06 & 14.26$\pm$0.06 & 14.29$\pm$0.28 & 13.57$\pm$0.25 &     &  &  &  & ex1,AGN? & \\ 
72  & \object{J162719.67-244148.9} &     & 18.68$\pm$0.03 & 16.02$\pm$0.05 & 15.15$\pm$0.05 & 14.57$\pm$0.05 & 13.65$\pm$0.08 & 13.6 $\pm$0.12 & 13.2 $\pm$0.28 &    &     &  &  &  &  & \\ 
73  & \object{J162721.11-243753.8} &     &    & 22.83$\pm$0.44 & 19.47$\pm$0.09 & 17.36$\pm$0.05 & 15.72$\pm$0.09 & 15.44$\pm$0.11 &    &    &     &  &  &  &  & \\ 
74  & \object{J162722.41-243837.9} &     &    & 22.2 $\pm$0.27 & 19.19$\pm$0.08 & 16.92$\pm$0.05 & 15.22$\pm$0.07 & 14.91$\pm$0.08 & 14.53$\pm$0.29 &    &     &  &  &  &  & \\ 
75  & \object{J162724.10-242510.9} &     &    & 22.32$\pm$0.7  & 18.73$\pm$0.07 & 15.85$\pm$0.05 & 13.77$\pm$0.06 & 13.22$\pm$0.06 & 12.8 $\pm$0.1  &    &     &  &  &  &  & \\ 
76  & \object{J162724.39-244147.8} &     &    & 18.78$\pm$0.05 & 15.01$\pm$0.05 & 12.72$\pm$0.05 & 10.93$\pm$0.06 & 10.47$\pm$0.06 & 10.0 $\pm$0.06 & 10.11$\pm$0.06 &     &  &  & 1,2,3,5,8  & & CRBR 2422.6-3507 \\ 
77  & \object{J162725.64-243728.6} &     & 22.22$\pm$0.03 & 18.23$\pm$0.05 & 16.54$\pm$0.05 & 15.36$\pm$0.05 & 14.16$\pm$0.06 & 13.9 $\pm$0.06 & 13.52$\pm$0.11 & 13.42$\pm$0.19 &     &  &  &  &  & \\ 
78  & \object{J162726.23-241923.1} &  23.56$\pm$0.03 & 20.46$\pm$0.03 & 16.35$\pm$0.05 & 14.36$\pm$0.05 & 13.04$\pm$0.05 & 11.64$\pm$0.06 & 11.14$\pm$0.05 & 10.62$\pm$0.06 & 9.93 $\pm$0.05 & 7.79$\pm$0.12  &  &  & 1,4,8 & var,ex1,ex2 & ISO-Oph 138 \\ 
79  & \object{J162726.61-244045.3} &     &    & 18.49$\pm$0.05 & 14.89$\pm$0.05 & 12.54$\pm$0.05 &    &    &    &    &     &  &  & 3,5,8 &  & 2MASS J16272661-2440451 \\ 
80  & \object{J162727.65-243827.2} &     &    & 21.94$\pm$0.21 & 18.94$\pm$0.07 & 16.95$\pm$0.05 & 15.3 $\pm$0.07 & 14.87$\pm$0.08 & 14.79$\pm$0.28 &    &     &  &  &  &  & \\ 
81  & \object{J162730.55-241456.5} &     &    & 21.75$\pm$0.21 & 18.73$\pm$0.07 & 16.92$\pm$0.05 & 15.56$\pm$0.08 & 15.35$\pm$0.09 &    &    &     &  &  &  &  & \\ 
82  & \object{J162730.69-244417.9} &     &    & 21.57$\pm$0.18 & 18.82$\pm$0.07 & 17.06$\pm$0.05 & 15.45$\pm$0.07 & 14.94$\pm$0.07 & 15.01$\pm$0.28 &    &     &  &  &  &  & \\ 
83  & \object{J162731.74-243148.7} &     &    & 21.92$\pm$0.24 & 18.47$\pm$0.06 & 16.44$\pm$0.05 & 15.05$\pm$0.06 & 14.66$\pm$0.07 & 14.57$\pm$0.3  &    &     &  &  &  &  & \\ 
84  & \object{J162732.13-242943.8} &     &    & 18.35$\pm$0.05 & 15.04$\pm$0.05 & 13.05$\pm$0.05 & 11.33$\pm$0.06 & 10.73$\pm$0.05 & 10.25$\pm$0.06 & 9.69 $\pm$0.06 & 6.22$\pm$0.12  &  &  & 5,8 & ex1,ex2 & BKLT J162732-242943 \\ 
85  & \object{J162732.55-241604.5} &     &    & 22.42$\pm$0.33 & 18.81$\pm$0.07 & 16.57$\pm$0.05 & 15.03$\pm$0.07 & 14.68$\pm$0.08 & 14.21$\pm$0.25 &    &     &  &  &  &  & \\ 
86  & \object{J162732.73-244500.5} &     &    & 19.03$\pm$0.05 & 15.41$\pm$0.05 & 13.25$\pm$0.05 & 11.63$\pm$0.06 & 11.18$\pm$0.06 & 10.91$\pm$0.05 & 10.79$\pm$0.06 & 8.9 $\pm$0.17  &  &  & 5.8 & ex2 & BKLT J162732-244500 \\ 
87  & \object{J162734.13-243308.9} &     &    & 22.22$\pm$0.26 & 18.96$\pm$0.07 & 17.03$\pm$0.05 & 15.53$\pm$0.07 & 15.23$\pm$0.07 & 14.53$\pm$0.24 &    &     &  &  &  &  & \\ 
88  & \object{J162734.47-241439.7} &     &    & 21.93$\pm$0.71 & 19.05$\pm$0.08 & 17.24$\pm$0.05 & 15.84$\pm$0.08 & 15.58$\pm$0.13 &    &    &     &  &  &  &  & \\ 
89  & \object{J162736.04-244325.4} &     &    & 22.84$\pm$0.49 & 19.05$\pm$0.08 & 16.71$\pm$0.05 & 15.1 $\pm$0.07 & 14.72$\pm$0.07 & 14.3 $\pm$0.18 &    &     &  &  &  &  & \\ 
90  & \object{J162736.59-245136.1} &     & 19.85$\pm$0.03 & 16.83$\pm$0.05 & 15.65$\pm$0.05 & 14.85$\pm$0.05 & 13.86$\pm$0.06 & 13.53$\pm$0.05 & 13.15$\pm$0.08 & 12.32$\pm$0.07 & 8.91$\pm$0.28  &  &  &  & ex1,ex2 & \\ 
91  & \object{J162737.03-244334.9} &     &    & 21.91$\pm$0.22 & 19.16$\pm$0.08 & 17.0 $\pm$0.05 & 15.62$\pm$0.08 & 15.26$\pm$0.11 &    &    &     &  &  &  &  & \\ 
92  & \object{J162737.21-243434.2} &     &    & 21.92$\pm$0.14 & 18.21$\pm$0.05 & 15.97$\pm$0.05 & 14.17$\pm$0.06 & 13.75$\pm$0.06 & 13.39$\pm$0.11 & 13.66$\pm$0.23 &     &  &  &  &  & \\ 
93  & \object{J162737.22-242526.7} &     &    & 22.31$\pm$0.36 & 18.81$\pm$0.07 & 16.18$\pm$0.05 & 14.43$\pm$0.06 & 13.97$\pm$0.06 & 13.73$\pm$0.13 &    &     &  &  &  &  & \\ 
94  & \object{J162738.95-244020.7} &  22.27$\pm$0.01 & 20.13$\pm$0.03 & 16.48$\pm$0.05 & 14.12$\pm$0.05 & 12.56$\pm$0.05 & 10.62$\pm$0.06 & 9.57 $\pm$0.05 & 8.7  $\pm$0.05 & 7.7  $\pm$0.05 & 2.98$\pm$0.1   &  &  & 1,3,4,8 & var,ex2,R & 2MASS J16273894-2440206 \\ 
95  & \object{J162740.12-242636.6} &     &    & 21.9 $\pm$0.19 & 17.39$\pm$0.05 & 14.13$\pm$0.05 & 11.62$\pm$0.06 & 10.93$\pm$0.05 & 10.38$\pm$0.05 & 10.03$\pm$0.06 & 7.67$\pm$0.12  &  &  & 4,5,8 & ex1,ex2 & GDS J162648.3-242834 \\ 
96  & \object{J162740.84-242900.8} &  18.96$\pm$0.01  & 18.7 $\pm$0.03 & 14.6 $\pm$0.05 & 13.76$\pm$0.05 & 13.19$\pm$0.05 & 12.51$\pm$0.06 & 12.24$\pm$0.06 & 11.79$\pm$0.06 & 10.96$\pm$0.06 & 8.33$\pm$0.17  &  M8.25$^{+0.25}_{-1.0}$ &    1.10$^{+0.7}_{-0.1}$    & 3,4,5,8 & ex1,ex2 & BKLT J162740-242901 \\ 
97  & \object{J162741.61-244644.8} &     & 20.32$\pm$0.03 & 17.87$\pm$0.05 & 15.87$\pm$0.05 & 14.22$\pm$0.05 & 11.32$\pm$0.05 & 10.28$\pm$0.05 & 9.38 $\pm$0.05 & 8.38 $\pm$0.05 & 5.04$\pm$0.1   &  &  & 1,8 & var,ex1,ex2 & BKLT J162740-242901 \\ 
98  & \object{J162744.20-235852.4} &     &    & 17.08$\pm$0.05 & 15.78$\pm$0.05 & 14.98$\pm$0.05 &    & 13.32$\pm$0.07 & 12.98$\pm$0.1  & 12.28$\pm$0.11 &     &  &  &  &  & \\ 
99  & \object{J162745.77-244453.8} &     & 22.08$\pm$0.03 & 17.74$\pm$0.05 & 14.8 $\pm$0.05 & 12.61$\pm$0.05 & 9.7  $\pm$0.06 & 8.71 $\pm$0.06 & 7.97 $\pm$0.05 & 7.27 $\pm$0.06 & 3.88$\pm$0.1   &  &  & 1,3,4,8  & ex2 & BKLT J162745-244454 \\ 
100 & \object{J162746.54-240559.2} &     & 21.51$\pm$0.03 & 17.94$\pm$0.05 & 16.34$\pm$0.05 & 15.26$\pm$0.05 & 14.41$\pm$0.07 & 14.17$\pm$0.08 & 13.96$\pm$0.15 &    &     &  &  &  &  & \\ 
101 & \object{J162747.25-244645.9} &     &    & 19.26$\pm$0.05 & 16.48$\pm$0.05 & 14.78$\pm$0.05 & 13.35$\pm$0.06 & 13.01$\pm$0.06 & 12.72$\pm$0.07 & 12.65$\pm$0.11 &     &  &  &  &  & \\ 
102 & \object{J162750.77-244245.0} &     &    & 22.87$\pm$0.41 & 19.42$\pm$0.08 & 17.18$\pm$0.05 & 15.6 $\pm$0.07 & 15.32$\pm$0.08 & 14.95$\pm$0.27 &    &     &  &  &  &  & \\ 
103 & \object{J162810.46-242420.4} &     & 23.49$\pm$0.05 & 17.74$\pm$0.05 & 16.15$\pm$0.05 & 15.07$\pm$0.05 & 13.98$\pm$0.06 & 13.68$\pm$0.06 & 13.29$\pm$0.1  & 12.97$\pm$0.12 &     &  &  &  & ex1 & \\ 
104 & \object{J162811.61-243729.9} &     &    & 19.51$\pm$0.05 & 16.95$\pm$0.05 & 15.41$\pm$0.05 & 13.81$\pm$0.06 & 13.22$\pm$0.06 & 12.62$\pm$0.07 & 11.92$\pm$0.07 &     &  &  &  & ex1 & \\ 
105 & \object{J162821.70-244247.3} &     & 21.86$\pm$0.03 & 16.66$\pm$0.05 & 14.15$\pm$0.05 & 12.48$\pm$0.05 & 10.84$\pm$0.05 & 10.24$\pm$0.05 & 9.7  $\pm$0.05 & 9.23 $\pm$0.05 & 6.18$\pm$0.1   &  &  & 1,8 & ex1,ex2 & BKLT J162821-244246 \\ 
106 & \object{J162829.93-245406.4} &     &    & 15.36$\pm$0.05 & 14.51$\pm$0.05 & 13.85$\pm$0.05 & 13.05$\pm$0.06 & 12.57$\pm$0.05 & 12.11$\pm$0.07 & 11.47$\pm$0.07 &     &  M6.50$^{+1.25}_{-1.0}$ &      4.9$^{+0.6}_{-0.7}$   & & ex1 &  \\ 
107 & \object{J162848.71-242631.8} &     &    & 14.31$\pm$0.05 & 13.6 $\pm$0.05 & 13.13$\pm$0.05 & 12.55$\pm$0.06 & 12.26$\pm$0.06 & 11.88$\pm$0.06 & 11.29$\pm$0.06 & 9.73$\pm$0.28  &  &  &  & ex1,ex2 & BKLT J162848-242631 \\ 
108 & \object{J162857.87-244055.1} &     &    & 18.74$\pm$0.05 & 15.93$\pm$0.05 & 14.01$\pm$0.05 & 11.15$\pm$0.06 & 10.06$\pm$0.05 & 9.13 $\pm$0.05 & 8.06 $\pm$0.05 & 4.12$\pm$0.1   &  &  &  & ex1,ex2 & SSTc2d J162857.9-244055 \\ 
109 & \object{J162902.95-244040.8} &     &    & 21.64$\pm$0.17 & 18.98$\pm$0.08 & 17.08$\pm$0.05 & 16.03$\pm$0.09 & 15.7 $\pm$0.11 &    &    &     &  &  &  &  & \\ 
110 & \object{J162903.97-244105.1} &     &    & 22.76$\pm$0.45 & 19.38$\pm$0.1  & 17.39$\pm$0.05 & 16.01$\pm$0.09 & 15.51$\pm$0.09 &    &    &     &  &  &  &  & \\ 
\footnotetext[1]{J2000.0 IAU designation.}
\footnotetext[2]{Aperture photometry magnitudes from PrimeCam / Subaru.}
\footnotetext[3]{PSF photometry magnitudes from WIRCam / CFHT.} 
\footnotetext[4]{IRAC and MIPS / Spitzer data as retrieved from the NASA/ IPAC Infrared Science Archive.}
\footnotetext[5]{Spectral Type and A$_{\emph{V}}$ as determined from this study. See Sect.~\ref{specfit} for details.}
\footnotetext[6]{{$\rho$~Ophiuchi members, according to the following studies: 1.~\citet{Bontemps2001}; 2.~\citet{Comeron1993}; 3.~\citet{Greene1992}; 4.~\citet{Gutermuth2009}; 5.~\citet{Strom1995}; 6.~\citet{Wilking1983}; 7.~\citet{Wilking1989}; 8.~\citet{Wilking2008}.} }
\footnotetext[7]{ex1, ex2~=~mid-IR excess;  Var~=~Variability; AGN?~=~Possible Active Galactic Nuclei Contaminant; R~=~red spectrum. See Sects.~\ref{discussion_phot}, and \ref{specfit} for details.} 
\footnotetext[8]{WL~20S, component of the triple system WL~20.}
\end{longtable}
\end{landscape}}

\end{document}